\documentclass[lettersize,journal]{IEEEtran}
\usepackage{amsmath,amsfonts}
\usepackage{algorithmic}
\usepackage{algorithm}
\usepackage{array}
\usepackage[caption=false,font=normalsize,labelfont=sf,textfont=sf]{subfig}
\usepackage{textcomp}
\usepackage{stfloats}
\usepackage{url}
\usepackage{verbatim}
\usepackage{graphicx}
\usepackage{cite}
\usepackage{tikz}
\usepackage[hidelinks]{hyperref}
\usepackage{orcidlink}
\usetikzlibrary{positioning}
\begin{document}

\title{MVLA-GR: A Phase-Free Multipath-Based Geometry Reconstruction Method via Multi-View Likelihood Accumulation for ISAC}

\author{Bowei~Xing\orcidlink{0009-0000-4405-0059},~\IEEEmembership{Graduate~Student~Member,~IEEE},
	Yuxiang~Zhang\orcidlink{0000-0003-0597-0594},~\IEEEmembership{Member,~IEEE},
	Jianhua~Zhang\orcidlink{0000-0002-6492-3846},~\IEEEmembership{Fellow,~IEEE},
	Yifeng~Xiong\orcidlink{0000-0002-4290-7116},~\IEEEmembership{Member,~IEEE},
	Hongbo~Xing\orcidlink{0009-0004-8889-4500},~\IEEEmembership{Graduate~Student~Member,~IEEE},
	Li~Yu\orcidlink{0000-0002-5782-1147},~\IEEEmembership{Member,~IEEE},
	and~Guangyi~Liu\orcidlink{0000-0002-8656-1946},~\IEEEmembership{Senior~Member,~IEEE}
	\thanks{This work was supported in part by National Natural Science Foundation of China (62401084, 62525101), National Key R\&D Program of China (2023YFB2904803, 2023YFB2904805), National Science and Technology Major Project (2025ZD1301700), Guangdong Major Project of Basic and Applied Basic Research (2023B0303000001), Natural Science Foundation of Beijing Xiaomi Innovation Joint Foundation (L243002), and the Beijing University of Posts and Telecommunications-China Mobile Communications Group Co., Ltd. Joint Institute. \textit{(Corresponding author: Jianhua Zhang.)}}
	\thanks{B. Xing, Y. Zhang, J. Zhang, H. Xing, and L. Yu are with the State Key Laboratory of Networking and Switching Technology, Beijing University of Posts and Telecommunications, Beijing 100876, China (e-mail: bwxing@bupt.edu.cn; zhangyx@bupt.edu.cn; jhzhang@bupt.edu.cn; hbxing@bupt.edu.cn; li.yu@bupt.edu.cn).}
	\thanks{Y. Xiong is with the School of Information and Electronic Engineering, Beijing University of Posts and Telecommunications, Beijing 100876, China (e-mail: yifengxiong@bupt.edu.cn).}
	\thanks{G. Liu is with the Future Research Laboratory, China Mobile Research Institute, Beijing 100053, China (e-mail: liuguangyi@chinamobile.com).}}

\markboth{IEEE Transactions on Wireless Communications}%
{Xing \MakeLowercase{\textit{et al.}}: MVLA-GR for Phase-Free Geometry Reconstruction in ISAC}


\maketitle

\begin{abstract}
	Integrated sensing and communication (ISAC) enables wireless systems to reuse communication signals for environmental sensing, where reconstructing the geometry of surrounding objects is a representative sensing task. However, many conventional methods rely on coherent processing and require accurate phase information, which is often hard to guarantee in practical communication systems, particularly at high carrier frequencies. To address this problem, this paper proposes a Multi-View Likelihood Accumulation Geometry Reconstruction (MVLA-GR) method based on channel impulse response (CIR) measurements, which uses only delay and power observations without requiring phase information. The method extracts dominant multipath components from each observation, and for each candidate spatial location, accumulates components across views whose propagation distances match the location as supporting evidence. A soft distance-matching kernel is introduced to tolerate range estimation errors and viewpoint-dependent scattering migration, and the received power of each component is used as a reliability weight. A joint thresholding strategy combining response magnitude and angular support continuity then converts the continuous support map into a binary geometry estimate. Ray-tracing simulations on canonical and complex targets, as well as real-world vehicle measurements at 36~GHz, demonstrate that MVLA-GR can effectively recover target geometry, providing a low-complexity phase-free solution for ISAC.
\end{abstract}

\begin{IEEEkeywords}
	Integrated sensing and communication (ISAC), environmental geometry reconstruction, channel impulse response (CIR), phase-free imaging, multi-view likelihood accumulation, multipath component extraction.
\end{IEEEkeywords}

\section{Introduction}
\label{section-1}
Integrated sensing and communication (ISAC) enables wireless systems to reuse communication signals for sensing the surrounding environment~\cite{liu2024cooperative,liu2022integrated,dong2023sensing}. Among various sensing tasks such as target detection, parameter estimation, and contour characterization~\cite{wang2024crb}, reconstructing the geometry of surrounding objects, such as their contour, shape, and spatial layout, is of particular interest because it provides a structural description of the propagation environment rather than isolated target parameters. Such geometric information is a key enabler for emerging concepts such as the digital twin channel, where a synchronized geometric model of the environment is used to characterize and predict wireless propagation~\cite{rekp,dtc_6g}. It also benefits various communication tasks in 6G networks, including environment-aware beamforming, blockage prediction, and proactive beam management.

In ISAC systems, pilot signals embedded in communication waveforms are commonly used for channel estimation, from which the channel impulse response (CIR) can be obtained as a basic observation. A wideband CIR describes the multipath structure of the wireless channel in the delay domain, where each propagation delay corresponds to a propagation distance and each multipath component reflects a scattering or reflection event in the environment, whose strength further depends on the radar cross section of the underlying scatterer~\cite{zhang2026rcs}. As a result, the CIR inherently carries geometric information about the surrounding scatterers~\cite{leitinger2019belief}, and its acquisition is naturally aligned with the communication-native operation of ISAC. This makes CIR a particularly suitable observation for environmental geometry reconstruction. Along this direction, the wireless environmental information theory has recently been proposed as a new paradigm for 6G environment intelligence communication~\cite{weit}.

A common way to reconstruct environmental geometry from wave-domain observations is to exploit the complex-valued wave response through coherent imaging. Representative examples include synthetic aperture radar (SAR) imaging~\cite{cumming2005digital}, microwave tomography~\cite{pastorino2010microwave}, and inverse-scattering reconstructions~\cite{jiang2024eps,jiang2025epsISAC,luo2026sgrm,lin2024envrec}, which jointly use the magnitude and phase of the measured field to reconstruct target shape or contrast. These methods can provide accurate reconstructions when reliable magnitude and phase measurements are available. However, in practical ISAC scenarios, although the magnitude of the CIR can be reliably extracted from communication pilots, the absolute phase reference of each CIR measurement and its consistency across observation positions remain challenging to maintain. Such challenges arise from two distinct sources. From the hardware perspective, oscillator drift, timing offsets, and the lack of cross-view phase synchronization at high carrier frequencies introduce per-observation phase offsets that are not easily predictable from one observation to the next~\cite{chung2022phasenoise,myers2019cfo}. From the physical perspective, the reflection phase introduced by surface scattering depends on both material composition and incidence geometry through the complex-valued Fresnel reflection coefficient, and varies non-trivially with viewpoint, especially for materials and surface conditions that are not known a priori. These factors together limit the practical applicability of coherent imaging in ISAC scenarios. From this perspective, phase-free approaches that rely on the magnitude information in the CIR are intrinsically robust to such phase uncertainties and are therefore better suited for ISAC scenarios.

Phase-free reconstruction has been studied along two main directions. One direction is phaseless inverse scattering, which reconstructs object shape or material distribution from the magnitude or intensity of scattered fields by solving a wave-equation-based nonlinear inverse problem~\cite{phaseless_isp1,phaseless_isp2,phaseless_dsm}. Such methods typically require a well-defined forward electromagnetic model and dense spatial or frequency sampling, and involve nonlinear iterative optimization, which makes them less suitable for lightweight CIR-based sensing in ISAC. The other direction is direct spatial accumulation, which projects measured delay-domain responses such as CIRs or PDPs back into the spatial domain and accumulates them over multiple observation positions~\cite{bp_radar,bp_gpr}. Incoherent backprojection is a representative example of this direction, and CIR-based heatmap construction follows a similar principle.

Among these two directions, direct spatial accumulation is more compatible with the lightweight, communication-native nature of ISAC, but it still has clear limitations when applied to CIR-based geometry reconstruction. First, direct spatial accumulation projects the entire PDP back into space, so that not only the dominant peaks but also sidelobes, noise floor, and other non-peak portions contribute to the spatial support. Since these non-peak portions of the PDP do not correspond to physical scattering events, their accumulation tends to blur the reconstruction and make the result sensitive to the detailed shape of the measured PDP. Second, direct spatial accumulation is essentially a heuristic projection scheme that does not explicitly model the underlying observation process, such as the statistical behavior of distance estimation, the viewpoint dependence of dominant scattering, or the angular continuity of true scattering structures. These unmodeled factors directly affect the reliability of the accumulated support, but cannot be addressed within the projection framework itself.

To address these gaps, a phase-free environmental geometry reconstruction method based on multi-view likelihood accumulation is proposed in this paper, with CIR-derived delay and power observations as the only input. The main contributions are summarized as follows.

\begin{itemize}
	\item A phase-free framework for environmental geometry reconstruction in ISAC is established, in which the problem is formulated as voxel-wise support estimation from CIR measurements, and only delay and power observations extracted from the CIR are used as the basic input.
	
	\item Within this framework, the proposed method takes individual multipath components as the basic processing units rather than the entire PDP, and converts each detected component into a distance-based geometric constraint through multi-view likelihood accumulation. A soft distance-matching kernel is introduced to tolerate range estimation errors and viewpoint-dependent scattering migration, the received power of each component serves as a reliability weight during accumulation, and a joint thresholding strategy combining response magnitude and angular support continuity is further applied to suppress artifacts and convert the continuous support map into a binary geometry estimate.
	
	\item The proposed method is validated through ray-tracing simulations on canonical and complex targets under various angular sampling intervals and SNR conditions, as well as real-world vehicle measurements at 36~GHz. Experimental results demonstrate that MVLA-GR can effectively recover target geometry and outperforms a standard incoherent backprojection baseline in the majority of tested conditions, confirming its practical applicability for ISAC scenarios.
\end{itemize}

The remainder of this paper is organized as follows. Section~\ref{section-2} introduces the system model and formulates the phase-free environmental geometry reconstruction problem from CIR measurements. Section~\ref{section-3} presents the proposed MVLA-GR method, including multi-view likelihood accumulation, parameter design, and binary reconstruction. Section~\ref{section-4} provides simulation and measurement results to evaluate the proposed method.

\section{System Model and Problem Formulation}
\label{section-2}

\begin{figure}[!t]
	\centering
	\includegraphics[width=\columnwidth]{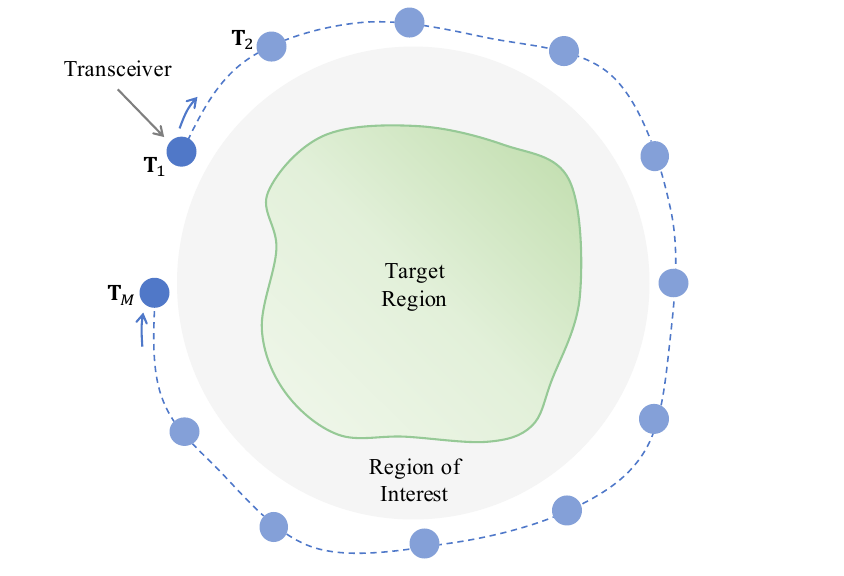}
	\caption{System model for multi-view geometry reconstruction. $\{\mathbf{T}_m\}$ denote the transceiver positions relative to the target.}
	\label{system_model_full}
\end{figure}

As illustrated in Fig.~\ref{system_model_full}, we consider a multi-view geometry reconstruction scenario in which a single omnidirectional transceiver collects echo signals at $M$ positions $\{\mathbf{T}_m\}$ relative to the target. As discussed in Section~\ref{section-4}, this multi-view observation structure can be physically realized either by a moving transceiver in a static environment, or by a stationary transceiver observing a moving target.

\subsection{Voxel Representation}

Under the Born approximation~\cite{born2013principles}, the electromagnetic interaction between the incident field and the target is treated as a single-scattering process in which each scattering element contributes independently, without accounting for mutual coupling or multiple reflections between elements. This assumption allows the target response to be represented by a set of spatially distributed scattering coefficients. Accordingly, the region of interest is discretized into $N$ voxels with centers $\mathbf{p}_n \in \mathbb{R}^3$, $n = 1,\ldots,N$, each associated with a nonnegative effective scattering intensity $x_n \geq 0$. The voxel intensity vector is defined as
\begin{equation}
	\mathbf{x} = [x_1, x_2, \ldots, x_N]^T \in \mathbb{R}_+^N.
	\label{eq:x}
\end{equation}

Here, $x_n$ is interpreted as an effective scattering intensity used for geometry reconstruction. Specifically, $x_n = 0$ indicates the absence of a scatterer, while $x_n > 0$ indicates the presence of a scatterer. The support of $\mathbf{x}$, i.e., the set of voxels for which $x_n > 0$, encodes the geometric structure and contour of the target. Therefore, the objective of the considered geometry reconstruction problem is not to recover the exact value of every $x_n$, but to determine which voxels satisfy $x_n > 0$.

\subsection{Electromagnetic Propagation and Received Signal Model}

To establish the relationship between the voxel scattering coefficients $\{x_n\}$ and the received signal at each observation position, we characterize the round-trip propagation between the transceiver and each voxel. Starting from the free-space Green's function
\begin{equation}
	G(\mathbf{r}, \mathbf{r}') = \frac{e^{\,jk\|\mathbf{r}-\mathbf{r}'\|}}{4\pi\|\mathbf{r}-\mathbf{r}'\|},
	\label{eq:green}
\end{equation}
where $k = 2\pi/\lambda$ is the wavenumber and $\lambda = c/f_c$ is the carrier wavelength, the round-trip propagation between the $m$-th transceiver position $\mathbf{T}_m$ and voxel $n$ is modeled as the product of two such Green's functions, leading to an effective complex channel gain
\begin{equation}
	g_{m,n} = \frac{e^{\,j\phi_{m,n}}}{(4\pi r_{m,n})^2},
	\label{eq:gain}
\end{equation}
where $r_{m,n} = \|\mathbf{p}_n - \mathbf{T}_m\|$ is the geometric distance between voxel $n$ and the $m$-th transceiver position. The phase term $\phi_{m,n}$ accounts for the round-trip propagation phase $2 k r_{m,n}$ together with hardware-induced contributions such as oscillator drift and timing offsets, which are not directly available in practice. In this model, $g_{m,n}$ captures the round-trip propagation effect, while $x_n$ represents the effective scattering intensity associated with voxel $n$.

A wideband pilot signal $s(t)$ of duration $T$ and bandwidth $B$ is assumed to be transmitted at each observation position and reused for sensing. The specific waveform of $s(t)$ is not restricted in this work; any standard wideband pilot whose autocorrelation provides a main-lobe range resolution on the order of $c/(2B)$ is applicable, such as OFDM training sequences, linear frequency-modulated chirps, or pseudo-random sequences commonly used in communication systems. The corresponding round-trip delay from the $m$-th transceiver position to voxel $n$ is
\begin{equation}
	\tau_{m,n} = \frac{2r_{m,n}}{c},
	\label{eq:delay}
\end{equation}
where $c$ denotes the speed of light. The received signal at the $m$-th transceiver position consists of target echoes, reflections from the surrounding environment, and additive noise:
\begin{equation}
	y_m(t) = \sum_{n=1}^{N} x_n\, g_{m,n}\, s(t - \tau_{m,n}) + s_{\mathrm{env},m}(t) + \eta_m(t),
	\label{eq:received}
\end{equation}
where $s_{\mathrm{env},m}(t)$ denotes reflections from objects outside the imaging region and $\eta_m(t)$ denotes the receiver noise.

Since the imaging region occupies a bounded spatial area, the target echoes correspond to a well-defined interval of round-trip delays determined by the geometry of the imaging region and the transceiver trajectory. Reflections from objects outside this region arrive at delays outside this interval and can be suppressed by range gating. After retaining only the delay interval $[\tau_{\min}, \tau_{\max}]$ corresponding to the imaging region, the received signal is modeled as
\begin{equation}
	y_m(t) = \sum_{n=1}^{N} x_n\, g_{m,n}\, s(t - \tau_{m,n}) + \eta_m(t).
	\label{eq:received_gated}
\end{equation}

\subsection{CIR Measurement Model}

The CIR is obtained by correlating the received signal $y_m(t)$ with the known pilot $s(t)$ over the observation interval of duration $T$:
\begin{equation}
	h_m(\tau) = \int_0^T y_m(t)\, s^*(t - \tau)\,\mathrm{d}t.
	\label{eq:CIR_cont}
\end{equation}
Substituting~\eqref{eq:received_gated} into~\eqref{eq:CIR_cont} yields
\begin{equation}
	h_m(\tau) = \sum_{n=1}^{N} x_n\, g_{m,n}\, p(\tau - \tau_{m,n}) + \tilde{\eta}_m(\tau),
	\label{eq:CIR}
\end{equation}
where
\begin{equation}
	p(\tau) = \int_0^T s(t)\,s^*(t-\tau)\,\mathrm{d}t
\end{equation}
is the autocorrelation of the pilot waveform, serving as the effective point-spread function of the system, and $\tilde{\eta}_m(\tau)$ is the filtered noise term. For a pilot signal occupying bandwidth $B$, the range resolution is
\begin{equation}
	\Delta R = \frac{c}{2B},
	\label{eq:resolution}
\end{equation}
which determines the minimum resolvable separation in the range dimension.

Discretizing the delay axis into $L$ taps $\{\tau_l = l/B\}_{l=0}^{L-1}$ and evaluating~\eqref{eq:CIR} at each tap, the CIR can be written in matrix-vector form as
\begin{equation}
	\mathbf{h}_m = \mathbf{G}_m \mathbf{x} + \boldsymbol{\eta}_m,
	\label{eq:forward}
\end{equation}
where $\mathbf{h}_m = [h_m(\tau_0),\ldots,h_m(\tau_{L-1})]^T \in \mathbb{C}^L$, $\boldsymbol{\eta}_m \in \mathbb{C}^L$ is the noise vector, and $\mathbf{G}_m \in \mathbb{C}^{L \times N}$ is the observation matrix with $(l,n)$-th element
\begin{equation}
	[\mathbf{G}_m]_{l,n} = g_{m,n}\, p(\tau_l - \tau_{m,n}).
	\label{eq:Gmln}
\end{equation}

In~\eqref{eq:Gmln}, the delay $\tau_{m,n}$ is geometrically determined by the known positions $\mathbf{T}_m$ and $\mathbf{p}_n$ via~\eqref{eq:delay}. However, fully specifying~\eqref{eq:Gmln} also requires phase information that is reliable and mutually consistent across different observation positions. In the considered mobile single-transceiver system, this condition is difficult to guarantee due to the high phase sensitivity at high frequencies, hardware-dependent phase offsets, and the lack of precise cross-view phase synchronization. Coherent inversion based on $\mathbf{G}_m$ is therefore not pursued in this work, and the subsequent development relies only on phase-insensitive features extracted from the CIR.

It is worth noting that, although the CIR $h_m(\tau)$ is itself complex-valued, the per-observation phase factors arising from oscillator drift, sampling timing offsets, and transceiver localization errors appear as global multiplicative unit-modulus factors on $h_m(\tau)$, which are eliminated by the squared-magnitude operation $|h_m(\tau)|^2$. The squared-magnitude operation therefore yields a stable representation that does not depend on these uncertain phase quantities, while remaining sensitive to the underlying multipath structure through the magnitude information.

Accordingly, instead of relying on phase, this work adopts the power delay profile (PDP), defined as
\begin{equation}
	P_m(\tau) = |h_m(\tau)|^2,
	\label{eq:PDP}
\end{equation}
as a phase-insensitive representation for subsequent geometry reconstruction.

\subsection{Phase-Free Observations}

Based on the phase-insensitive representation in~\eqref{eq:PDP}, this work extracts phase-free observations from the PDP at each observation position. Specifically, a standard peak detection procedure is applied to $P_m(\tau)$ to identify $K_m$ detected multipath components, each characterized by its detected delay and received power. The detection is based on local maxima of $P_m(\tau)$, with a noise-relative magnitude threshold to suppress spurious detections, and a minimum inter-peak delay separation on the order of $1/B$ to avoid splitting a single main lobe into multiple peaks.

Due to the finite range resolution $\Delta R$, multiple voxel contributions with sufficiently close round-trip delays may be mapped to the same detected multipath component in the PDP. Let $\mathcal{N}_{m,i}$ denote the index set of voxels contributing to the $i$-th detected component at the $m$-th view. Its complex amplitude can be written as
\begin{equation}
	\alpha_{m,i} = \sum_{n \in \mathcal{N}_{m,i}} x_n\, g_{m,n}\, p(\tau_{m,i} - \tau_{m,n}) + \tilde{\eta}_{m,i},
	\label{eq:alpha}
\end{equation}
where $\tilde{\eta}_{m,i} \in \mathbb{C}$ is the effective complex noise term at the corresponding delay bin after matched filtering. From each detected multipath component, two phase-free quantities are extracted:
\begin{align}
	R_{m,i} &= \frac{c\,\tau_{m,i}}{2},
	\label{eq:Rmi} \\
	p_{m,i} &= |\alpha_{m,i}|^2
	\approx \left|\sum_{n \in \mathcal{N}_{m,i}} x_n\, g_{m,n}\, p(\tau_{m,i} - \tau_{m,n})\right|^2
	+ \xi_{m,i},
	\label{eq:pmi_full}
\end{align}
where $R_{m,i}$ is the propagation distance inferred from the detected delay $\tau_{m,i}$, $p_{m,i}$ is the received power of the $i$-th detected multipath component, and $\xi_{m,i}$ denotes a power-domain perturbation arising from receiver noise and its cross-coupling with the signal term. The complete phase-free observation set at the $m$-th view is then defined as
\begin{equation}
	\mathcal{C}_m = \bigl\{(R_{m,i},\; p_{m,i})\bigr\}_{i=1}^{K_m},
	\label{eq:obs}
\end{equation}
where $K_m$ is the number of detected multipath components.

In the subsequent development, the distances $\{R_{m,i}\}$ provide the geometric constraints used for voxel localization, while the powers $\{p_{m,i}\}$ are used as reliability indicators for multi-view fusion.

\subsection{Problem Formulation}

Since the geometry reconstruction objective is to identify the support of $\mathbf{x}$, we introduce the voxel occupancy variable
\begin{equation}
	z_n =
	\begin{cases}
		1, & x_n > 0, \\
		0, & x_n = 0,
	\end{cases}
	\qquad n=1,\ldots,N,
	\label{eq:zn}
\end{equation}
and define the binary occupancy vector
\begin{equation}
	\mathbf{z} = [z_1,\ldots,z_N]^T \in \{0,1\}^N.
	\label{eq:z}
\end{equation}
The geometry reconstruction problem is therefore formulated as the estimation of the voxel occupancy pattern from the multi-view phase-free observations $\{\mathcal{C}_m\}_{m=1}^{M}$:
\begin{equation}
	\hat{\mathbf{z}}
	=
	\arg\max_{\mathbf{z}\in\{0,1\}^N}
	P\!\left(\mathbf{z}\mid \{\mathcal{C}_m\}_{m=1}^{M}\right).
	\label{eq:MAP_support}
\end{equation}

Solving~\eqref{eq:MAP_support} exactly requires the likelihood $P(\{\mathcal{C}_m\}_{m=1}^{M}\mid\mathbf{z})$, which in turn requires an explicit model relating the observed powers $\{p_{m,i}\}$ to the underlying voxel occupancies. However, as seen from~\eqref{eq:pmi_full}, the received power depends on the unknown phases embedded in $\{g_{m,n}\}$. Even if the occupied voxels were known, the observed powers could not be predicted explicitly without these phases. Consequently, the exact likelihood cannot be written in closed form, and the posterior in~\eqref{eq:MAP_support} is not directly tractable.

Nevertheless, the phase-free observations still contain useful information for support identification. In particular, the distance measurements $\{R_{m,i}\}$ provide geometric constraints on the possible locations of scatterers, while the received powers $\{p_{m,i}\}$ reflect the relative reliability of the detected multipath components. This suggests that the voxel occupancy can be approximated from multi-view geometric consistency, with received power serving as a reliability weight. The proposed MVLA-GR method in Section~\ref{section-3} is developed along this line by constructing a tractable approximation of the posterior probability of voxel occupancy from the observation sets $\{\mathcal{C}_m\}_{m=1}^{M}$.

\section{Multi-View Likelihood Accumulation Method for Geometry Reconstruction}
\label{section-3}

\begin{figure*}[t]
	\centering
	\includegraphics[width=\textwidth]{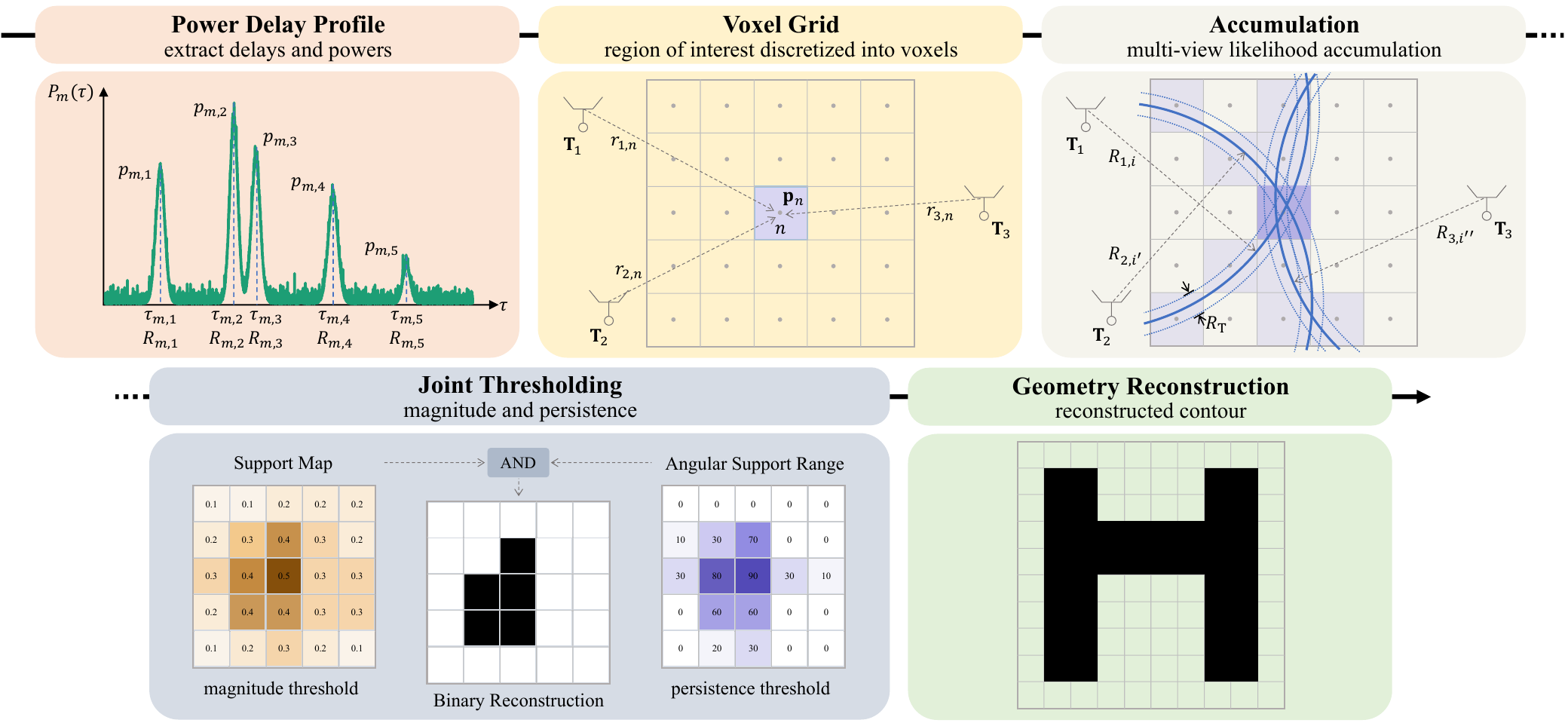}
	\caption{Overview of the proposed Multi-View Likelihood Accumulation Geometry Reconstruction (MVLA-GR) pipeline.}
	\label{fig:pipeline}
\end{figure*}

This section presents the proposed Multi-View Likelihood Accumulation Geometry Reconstruction (MVLA-GR) method. As discussed at the end of Section~\ref{section-2}, the posterior in~\eqref{eq:MAP_support} is not directly tractable, but the phase-free observations still carry useful geometric information. The proposed method exploits this information by approximating the posterior probability of voxel occupancy through two phase-free quantities that the CIR observations provide reliably, namely the propagation distances $\{R_{m,i}\}$ and the received powers $\{p_{m,i}\}$. The distances are used to test whether each voxel is geometrically consistent with the observed multipath components across views, and the powers are used to weight the contribution of each component according to its reliability. By accumulating these weighted geometric matches over all observation views, a voxel-wise support score $s_n$ is obtained as a tractable surrogate for the posterior probability of voxel occupancy. The support map $\{s_n\}_{n=1}^{N}$ is then converted into the final binary reconstruction $\hat{\mathbf{z}}$ through a joint thresholding step that combines response magnitude and angular support continuity.

These steps are summarized in the processing pipeline shown in Fig.~\ref{fig:pipeline}. From left to right, multipath components are first extracted from the PDP at each observation position to form the phase-free observations. The region of interest is then discretized into a voxel grid, and the multi-view geometric consistency between each voxel and the extracted distance--power observations is evaluated and accumulated into a support map. Finally, the joint thresholding step is applied to convert the continuous support map into a binary geometry estimate. The remainder of this section describes each of these components in detail.

\subsection{Multi-View Likelihood Accumulation}
\label{section-3a}

\subsubsection{Hard-Decision Occupancy Support}

For voxel $n$ and observation view $m$, consider the event that at least one detected propagation distance is geometrically consistent with the voxel distance, i.e.,
\begin{equation}
	\mathcal{E}_{m,n}
	=
	\left\{
	\exists\, i \in \{1,\dots,K_m\}
	\ \text{s.t.}\
	|R_{m,i}-r_{m,n}| \le \epsilon_R
	\right\},
	\label{eq:event_hard}
\end{equation}
where $R_{m,i}$ is the $i$-th detected propagation distance at view $m$, $r_{m,n}$ is the geometric distance between voxel $n$ and the $m$-th transceiver position, and $\epsilon_R$ is a distance tolerance related to the range resolution.

If voxel $n$ corresponds to a true scattering location, the event $\mathcal{E}_{m,n}$ is more likely to occur. In contrast, for an unoccupied voxel, this event can only occur accidentally with a much smaller probability. Based on this observation, we define the hard-decision support indicator
\begin{equation}
	I_{m,n} = \mathbb{I}(\mathcal{E}_{m,n}),
	\label{eq:hard_indicator}
\end{equation}
where $\mathbb{I}(\cdot)$ denotes the indicator function.

Treating the $M$ observation views as independent binary trials, the fraction of views in which the event $\mathcal{E}_{m,n}$ occurs provides a coarse approximation to the posterior probability of voxel occupancy:
\begin{equation}
	\tilde{P}_{\mathrm{hard}}\!\left(z_n=1 \mid \{\mathcal{C}_m\}_{m=1}^{M}\right)
	\approx
	\frac{1}{M}\sum_{m=1}^{M} I_{m,n},
	\label{eq:hard_posterior}
\end{equation}
where $\tilde{P}_{\mathrm{hard}}(\cdot)$ denotes a surrogate posterior probability rather than an exact one. Under this hard-decision model, voxels supported by a larger fraction of observation views are more likely to be occupied.

\subsubsection{Soft Distance-Matching Kernel}

The hard-decision rule in~\eqref{eq:hard_posterior} is not sufficiently robust in practice. First, the extracted distances $R_{m,i}$ are affected by finite bandwidth, peak detection errors, and measurement noise. Second, under viewpoint variation, the dominant scattering point associated with the same physical structure may migrate along the target surface, especially in specular-dominated scenarios. As a result, a strict binary decision may reject physically meaningful observations that exhibit only a small distance deviation.

To address these issues, we replace the hard-decision indicator by a soft probabilistic kernel derived from the statistics of the distance estimation error.

For each detected multipath component, we assume that its corresponding peak in the power delay profile is sufficiently isolated from adjacent components, and that the measurement SNR is sufficiently high. Under these conditions, the extracted propagation distance can be modeled as
\begin{equation}
	R_{m,i} = r^{\star}_{m,i} + \varepsilon_{m,i},
	\label{eq:range_error_model}
\end{equation}
where $R_{m,i}$ is the extracted propagation distance, $r^{\star}_{m,i}$ is the true propagation distance of the $i$-th component, and $\varepsilon_{m,i}$ is the associated distance estimation error. We assume that $\varepsilon_{m,i}$ approximately follows a zero-mean Gaussian distribution, i.e., $\varepsilon_{m,i} \sim \mathcal{N}(0,\sigma_\varepsilon^2)$. This assumption is motivated by the asymptotic normality of maximum-likelihood estimators under additive Gaussian noise~\cite{trees2001,kay1993}, and is commonly adopted as a tractable approximation in time-delay estimation problems~\cite{quazi1981}. In practical multipath scenarios where peaks are not fully isolated, this Gaussian model is still used here as an approximate error model for soft geometric matching.

Under this Gaussian error model, the conditional likelihood of observing distance $R_{m,i}$ given that voxel $n$ is occupied and located at geometric distance $r_{m,n}$ is proportional to
\begin{equation}
	P(R_{m,i} \mid z_n = 1,\, r_{m,n})
	\propto
	\exp\!\left(-\frac{(r_{m,n} - R_{m,i})^2}{2\sigma_R^2}\right),
\end{equation}
where $\sigma_R = \sigma_\varepsilon$ is the standard deviation of the range estimation error. Defining the distance mismatch
\begin{equation}
	d_{m,i}^{(n)} = r_{m,n} - R_{m,i},
	\label{eq:distance_mismatch_new}
\end{equation}
this conditional likelihood motivates the following truncated Gaussian kernel as the soft replacement for the hard indicator $I_{m,n}$:
\begin{equation}
	\phi\!\left(d_{m,i}^{(n)}\right)=
	\begin{cases}
		\exp\!\left(-\dfrac{(d_{m,i}^{(n)})^2}{2\sigma_R^2}\right),
		& |d_{m,i}^{(n)}| \le R_{\mathrm{T}}, \\[6pt]
		0, & \text{otherwise},
	\end{cases}
	\label{eq:likelihood_kernel_new}
\end{equation}
where $R_{\mathrm{T}}$ is a truncation radius beyond which the voxel is regarded as geometrically incompatible with the observation. The truncation ensures that clearly inconsistent distance observations do not contribute to the voxel support and also improves computational efficiency.

\subsubsection{Power-Based Weighting}
\label{section-3a3}

Different detected multipath components do not provide equally reliable geometric evidence and should therefore not contribute equally to voxel occupancy support. The received power $p_{m,i}$ is adopted as a reliability weight for two main reasons.

First, under matched filtering, the delay estimation error $\varepsilon_{m,i}$ decreases with increasing SNR. Components with higher received power generally correspond to higher effective SNR, and their extracted distances are therefore more reliable. Weighting by power naturally suppresses weak and noisy components whose geometric constraints are less trustworthy.

Second, the observed power also reflects the strength of the corresponding propagation path. Stronger detected components usually carry more informative evidence about the underlying scattering structure, while weaker components are more likely to be unstable or noise-contaminated. Therefore, assigning larger weights to stronger components helps the multi-view accumulation focus on more credible geometric evidence.

Accordingly, to preserve the relative reliability differences among components within the same view while preventing any single view from dominating the overall fusion, the weight for each detected component is defined through view-wise normalization:
\begin{equation}
	w_{m,i}
	=
	\frac{p_{m,i}}{\sum_{\ell=1}^{K_m}p_{m,\ell}}.
	\label{eq:nor_w_new}
\end{equation}
The view-wise normalization in~\eqref{eq:nor_w_new} implicitly assumes that the observation views are of comparable quality, which is typically valid in the considered multi-view sensing settings, where all observations are collected by the same transceiver in a controlled measurement campaign and therefore share similar noise levels and detection conditions. Under this assumption, the relative powers within each view directly reflect the relative reliability of the detected components.

\subsubsection{Accumulation and Posterior Approximation}

With the soft kernel and power-based weights defined above, the contribution from the $i$-th detected component at view $m$ to voxel $n$ is
\begin{equation}
	\ell_{m,i}^{(n)}
	=
	w_{m,i}\,
	\phi\!\left(d_{m,i}^{(n)}\right).
	\label{eq:single_likelihood_contribution}
\end{equation}

Accumulating all detected components within view $m$, we define the total support of voxel $n$ under the $m$-th observation as
\begin{equation}
	a_{m,n}
	=
	\sum_{i=1}^{K_m}
	w_{m,i}\,
	\phi\!\left(d_{m,i}^{(n)}\right),
	\label{eq:view_support_new}
\end{equation}
where $K_m$ is the number of detected multipath components at view $m$.

The quantity $a_{m,n}$ can be interpreted as a surrogate of the posterior probability of voxel occupancy under the $m$-th observation, i.e.,
\begin{equation}
	a_{m,n}
	\approx
	\tilde{P}_m\!\left(z_n=1 \mid \mathcal{C}_m\right),
	\label{eq:view_support_prob_interp}
\end{equation}
where $\tilde{P}_m(\cdot)$ denotes a surrogate posterior probability rather than an exact one. In this sense, $a_{m,n}$ is the soft generalization of the hard indicator $I_{m,n}$ in~\eqref{eq:hard_indicator}.

To aggregate evidence across all observation views, we define the normalized multi-view support score of voxel $n$ as
\begin{equation}
	s_n
	=
	\frac{1}{M}\sum_{m=1}^{M} a_{m,n}
	=
	\frac{1}{M}\sum_{m=1}^{M}\sum_{i=1}^{K_m}
	w_{m,i}\,
	\phi\!\left(d_{m,i}^{(n)}\right).
	\label{eq:voxel_accumulation_expanded}
\end{equation}
Accordingly, $s_n$ can be interpreted as a tractable approximation to the posterior probability of occupancy for voxel $n$ under the full multi-view observations, i.e.,
\begin{equation}
	s_n
	\approx
	\tilde{P}\!\left(z_n=1 \mid \{\mathcal{C}_m\}_{m=1}^{M}\right).
	\label{eq:sn_prob_interp}
\end{equation}
A voxel with a larger $s_n$ is supported by more observation views and more reliable multipath components, and is therefore more likely to be occupied. The collection $\{s_n\}_{n=1}^{N}$ forms the reconstructed occupancy map used in all subsequent processing.

The computational complexity of the proposed accumulation process is dominated by the evaluation of voxel-wise contributions for all detected multipath components across all observation views, resulting in
\begin{equation}
	\mathcal{O}\!\left(
	N\sum_{m=1}^{M}K_m
	\right)
	=
	\mathcal{O}(MN\bar K),
	\label{eq:complexity_new}
\end{equation}
where $N$ is the number of voxels, $K_m$ is the number of detected multipath components at view $m$, and $\bar K$ denotes the average number of components per view.

\subsection{Parameter Selection and Binary Reconstruction}
\label{section-3b}

The accumulation process in Section~\ref{section-3a} yields a continuous voxel-wise support score $s_n$, which serves as a surrogate of the posterior probability of voxel occupancy. To obtain a reliable binary reconstruction from this continuous support map, two issues must be further addressed. First, the kernel parameters should be chosen such that valid supports can accumulate across views while maintaining sufficient radial discrimination. Second, accidental overlaps among supports from unrelated scattering structures must be suppressed. These two aspects are discussed in the following subsections.

\subsubsection{Parameter $\sigma_R$ (Radial Kernel Scale)}
\label{section-3b1}

\begin{figure}[!t]
	\centering
	\includegraphics[width=\columnwidth]{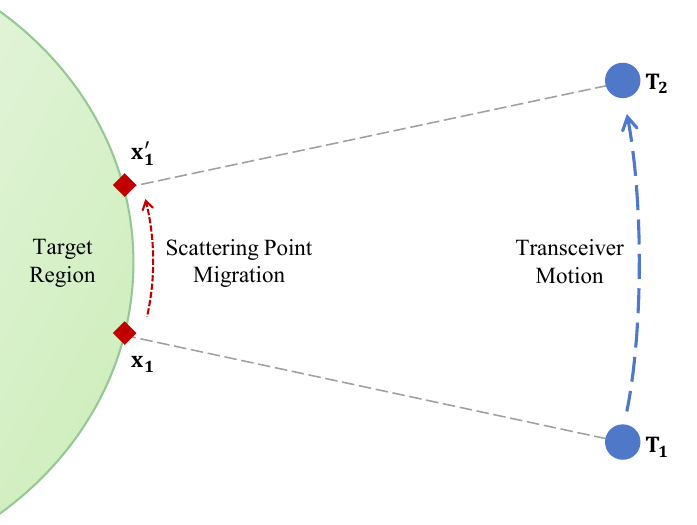}
	\caption{Illustration of viewpoint-dependent migration of dominant scattering locations.}
	\label{migrate}
\end{figure}

In~\eqref{eq:likelihood_kernel_new}, the parameter $\sigma_R$ controls how fast the soft support decays as the distance mismatch $d_{m,i}^{(n)}$ increases. As established above, $\sigma_R$ should reflect the effective uncertainty of the extracted propagation distance. Although this uncertainty in general depends on multiple factors, including receiver noise, the dominant scale in multipath-rich scenarios is set by the system range resolution $\Delta R = c/(2B)$, which determines the main-lobe width of the pulse shaping function $p(\tau)$ and thereby the achievable precision of peak detection. Accordingly, $\sigma_R$ is parameterized as
\begin{equation}
	\sigma_R = \kappa\,\Delta R,
	\label{eq:sigmaR_general}
\end{equation}
where $\kappa$ is a dimensionless proportionality constant controlling the tolerance of the radial kernel. A smaller $\kappa$ yields sharper radial discrimination but weaker robustness to range errors, whereas a larger $\kappa$ increases tolerance at the cost of reduced spatial selectivity. In practice, $\kappa$ should be chosen according to the trade-off between range uncertainty and discrimination ability, and is determined in the experimental section.

\subsubsection{Parameter $R_{\mathrm{T}}$ (Truncation Radius)}
\label{section-3b2}

The parameter $R_{\mathrm{T}}$ in~\eqref{eq:likelihood_kernel_new} determines the effective support radius of the radial kernel. Under a purely range-limited model, $R_{\mathrm{T}}$ would naturally be chosen on the same order as $\sigma_R$. In practical geometry reconstruction scenarios, however, the dominant scattering point associated with the same physical reflecting region may change with the observation viewpoint, especially under specular-dominated scattering conditions. As illustrated in Fig.~\ref{migrate}, when the transceiver moves from $\mathbf{T}_1$ to $\mathbf{T}_2$, the dominant scattering response may shift from one point on the target surface to another nearby point. Although these responses still originate from the same physical reflecting region, their effective propagation distances may differ due to viewpoint-dependent migration of the dominant scattering location.

If $R_{\mathrm{T}}$ is chosen solely according to the range-resolution scale, valid supports associated with the same physical structure may fail to overlap across different views, weakening the subsequent likelihood accumulation. To preserve cross-view overlap among supports corresponding to the same physical scattering region, the truncation radius should also account for viewpoint-induced scattering migration.

To this end, we introduce a migration scale
\begin{equation}
	\delta_{\mathrm{move}}=\eta D,
	\label{eq:delta_move_scale}
\end{equation}
where $D$ denotes the maximum spatial extent of the target and $\eta$ is a small dimensionless coefficient. The parameter $\delta_{\mathrm{move}}$ characterizes the typical range variation caused by viewpoint-dependent movement of dominant scattering locations. Its value depends on target geometry, surface properties, and angular sampling density, and should be selected to provide sufficient overlap among valid supports across views without excessively enlarging the kernel support.

\begin{figure}[!b]
	\centering
	\includegraphics[width=\columnwidth]{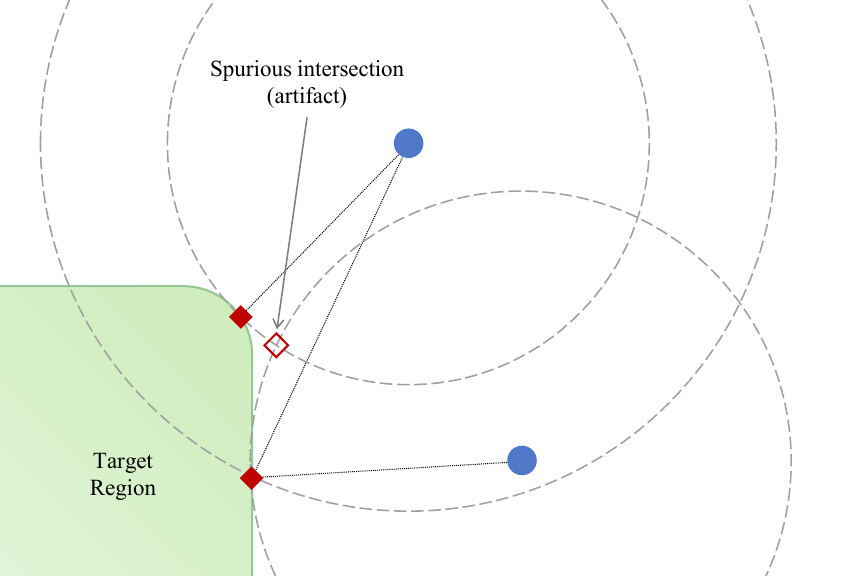}
	\caption{Illustration of reconstruction artifacts caused by accidental intersections of range constraints associated with different scattering structures.}
	\label{artifacts}
\end{figure}

Since the effective support radius must cover both the range-uncertainty scale and the migration scale, the truncation radius is chosen as
\begin{equation}
	R_{\mathrm{T}} = \max(\sigma_R,\delta_{\mathrm{move}}).
	\label{eq:RT_def_new}
\end{equation}
This ensures that the likelihood kernel remains compact enough to suppress clearly incompatible voxels, while still allowing valid supports from different observation views to overlap and accumulate on the same physical scattering structure.

\subsubsection{Joint Thresholding for Binary Reconstruction}
\label{section-3b3}

The accumulated support map $\{s_n\}_{n=1}^{N}$ provides a continuous spatial representation of voxel occupancy support. However, before converting this support map into a binary reconstruction, spurious responses caused by accidental overlaps among unrelated scattering structures must be suppressed. Such responses appear as reconstruction artifacts, as illustrated in Fig.~\ref{artifacts}.

A true scattering voxel is expected to be supported not only by a large accumulated response, but also over a sufficiently wide \emph{continuous} range of observation directions. By contrast, an artifact voxel is usually produced by accidental overlaps among unrelated observations and tends to be supported only within a limited or fragmented angular region. Therefore, in addition to the accumulated response magnitude, the continuity of voxel support across observation directions provides an important cue for artifact suppression.

To characterize this persistence, we first define a binary support indicator for voxel $n$ at view $m$ as
\begin{equation}
	b_{m,n}=
	\begin{cases}
		1, & a_{m,n}>0,\\
		0, & a_{m,n}=0,
	\end{cases}
\end{equation}
where $a_{m,n}>0$ holds if and only if at least one detected component at view $m$ satisfies $|d_{m,i}^{(n)}| \leq R_{\mathrm{T}}$, as defined in~\eqref{eq:likelihood_kernel_new}.

Let $\phi_m$ denote the observation angle of the $m$-th transceiver position with respect to the target center, and assume that $\{\phi_m\}$ are ordered increasingly. For voxel $n$, the support sequence $\{b_{m,n}\}_{m=1}^{M}$ generally forms several continuous supporting intervals along the angular axis. Denote these intervals by $\Omega_{n,\ell}$, $\ell=1,\dots,L_n$. For each interval $\Omega_{n,\ell}$, its angular width is defined as
\begin{equation}
	\Theta_{n,\ell}
	=
	\max_{m\in\Omega_{n,\ell}} \phi_m
	-
	\min_{m\in\Omega_{n,\ell}} \phi_m.
\end{equation}
The angular support range of voxel $n$ is then defined as the maximum width among all continuous supporting intervals:
\begin{equation}
	\Theta_n
	=
	\max_{\ell=1,\dots,L_n}\Theta_{n,\ell}.
	\label{eq:angular_support_range}
\end{equation}

To exploit both the response magnitude and the angular persistence, we adopt a joint thresholding strategy. First, a magnitude threshold is applied to the accumulated voxel responses $\{s_n\}$, and only voxels whose response values lie within the top $p\%$ of all voxels are retained. Second, an angular-support threshold $T_\theta$ is imposed, and only voxels satisfying
\begin{equation}
	\Theta_n \ge T_\theta
\end{equation}
are preserved. The joint thresholding criterion is therefore written as
\begin{equation}
	\text{voxel } n \text{ is retained if}
	\quad
	\Big(
	s_n \in \text{Top-}p\%,\;
	\Theta_n \ge T_\theta
	\Big).
	\label{eq:joint_thresholding_new}
\end{equation}

This criterion has a clear physical interpretation: a voxel is regarded as a reliable scattering location only if it exhibits both a high support score and a sufficiently wide continuous support range across observation directions. As a result, spurious voxels caused by accidental support overlap can be effectively suppressed, while voxels associated with true scattering structures are more likely to be retained.

The final binary reconstruction is then obtained by setting $\hat z_n = 1$ for retained voxels and $\hat z_n = 0$ otherwise, thereby yielding an estimate $\hat{\mathbf z}$ of the occupancy vector in~\eqref{eq:MAP_support}.

\section{Simulation and Measurement Results}
\label{section-4}

Before presenting the experimental results, we briefly note that the proposed framework applies to two complementary ISAC scenarios that share the same multi-view CIR observation structure: (i) a mobile transceiver collecting CIRs along its trajectory to reconstruct static surroundings, supporting tasks such as environment-aware beam management and digital twin construction; and (ii) a static base station observing a moving target whose successive positions provide the multi-view observations, supporting tasks such as target classification and behavior recognition. The simulations in this section follow the first scenario with a fixed circular trajectory, while the vehicle measurement in Section~\ref{subsec:measurement_validation} corresponds to the second scenario, where the antenna is fixed and the vehicle rotates on a turntable to emulate the relative aspect-angle variation between a stationary base station and a moving target.

The proposed MVLA-GR method is evaluated as follows. First, the model parameters are determined using several canonical geometric targets, so that all subsequent evaluations are carried out under a unified parameter setting rather than target-specific tuning. Second, under this fixed parameter configuration, the proposed method is quantitatively compared with a baseline method on a more complex star-shaped target under different angular sampling intervals and signal-to-noise ratios (SNRs). Finally, real-world measurement results on a vehicle target are presented to demonstrate the practical applicability of the proposed framework.

All simulation results in this section are generated using ray-tracing data produced by Remcom Wireless InSite. The carrier frequency is set to 36~GHz, and the signal bandwidth is 3~GHz, corresponding to a range resolution of $\Delta R = c/(2B) = 0.05$~m.

Unless otherwise specified, the transceiver positions are distributed along a circular trajectory with a radius of 10~m around the target, and the reconstruction region is discretized into a uniform voxel grid. It should be emphasized that the proposed MVLA-GR method does not rely on a circular trajectory itself, but only on the knowledge of the transceiver positions at which the CIR measurements are collected. The circular observation pattern adopted here is introduced purely as a convenient and reproducible simulation configuration.

In all simulation experiments, additive noise is introduced in the PDP domain after the clean CIR data are converted into PDPs. Multipath components are then re-extracted from the noisy PDPs by thresholding and peak detection, and the resulting phase-free observations are used for geometry reconstruction. The reconstruction performance in this section is evaluated using the Chamfer distance (CD). Let $\mathcal{X}=\{\mathbf{x}_u\}_{u=1}^{N_X}$ denote the set of ground-truth occupied voxel centers, and let $\mathcal{Y}=\{\mathbf{y}_v\}_{v=1}^{N_Y}$ denote the set of reconstructed occupied voxel centers. The Chamfer distance is defined as
\begin{equation}
	\begin{aligned}
		\mathrm{CD}
		=
		10\log_{10}\!\Bigg(
		&\frac{1}{N_X}\sum_{u=1}^{N_X}\min_{v}\|\mathbf{x}_u-\mathbf{y}_v\|_2^2 \\
		&\quad+
		\frac{1}{N_Y}\sum_{v=1}^{N_Y}\min_{u}\|\mathbf{y}_v-\mathbf{x}_u\|_2^2
		\Bigg).
	\end{aligned}
	\label{eq:chamfer_ch4}
\end{equation}
The resulting CD is expressed in dB(m$^2$), and a smaller CD indicates a closer geometric match between the reconstructed result and the ground-truth geometry.

\subsection{Parameter Determination Using Canonical Geometric Targets}
\label{subsec:param_selection}

Before conducting comparative evaluations on complex targets, the two model parameters in MVLA-GR, namely the kernel-scale parameter $\kappa$ and the migration parameter $\eta$, are first determined using several canonical geometric targets. This step is introduced to obtain a unified parameter setting from simple geometries, rather than tuning the method specifically for the star-shaped target used in the subsequent comparisons.

\subsubsection{Canonical Target Setup}

Three canonical targets are considered for parameter determination: a circle, a square, and an equilateral triangle. Their geometric sizes are chosen as follows: the circle has a radius of 4~m, the square has a side length of 8~m, and the equilateral triangle has a height of 6~m. These three targets represent smooth curved boundaries, straight edges, and sharp-corner structures, respectively, which correspond to three fundamental scattering mechanisms relevant to MVLA-GR: continuous scattering migration along curved surfaces, piecewise-constant reflection along flat segments, and edge diffraction at sharp corners. Since most practical targets are composed of these three basic scattering elements, parameters determined on this basis are expected to generalize across diverse target geometries. For the circle and square, the maximum extent $D$ is 8~m. For the equilateral triangle, the maximum extent $D$ is taken as its side length, which is $4\sqrt{3}$~m.

In this experiment, each target is observed from 72 uniformly distributed views over the full $360^\circ$ angular range, corresponding to an angular interval of $5^\circ$. The SNR is fixed to 20~dB, and the remaining system parameters are kept identical across the three targets.

\subsubsection{Selection Criterion for $\kappa$ and $\eta$}

\begin{figure}[!t]
	\centering
	\subfloat[]{
		\includegraphics[width=0.23\textwidth]{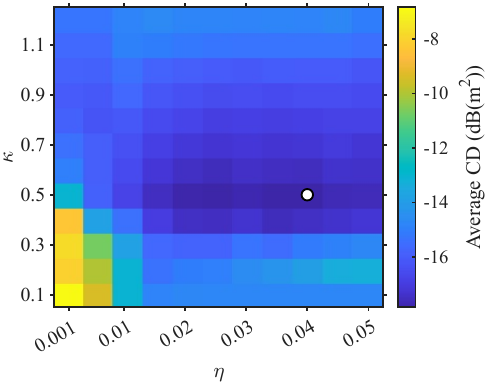}
		\label{fig:kappa_eta_sweep_a}}
	\hfill
	\subfloat[]{
		\includegraphics[width=0.23\textwidth]{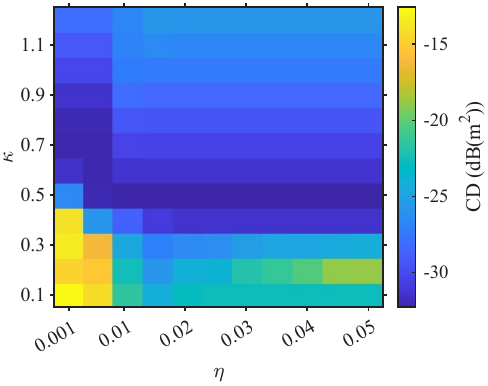}
		\label{fig:kappa_eta_sweep_b}}
	
	\vspace{0.1mm}
	
	\subfloat[]{
		\includegraphics[width=0.23\textwidth]{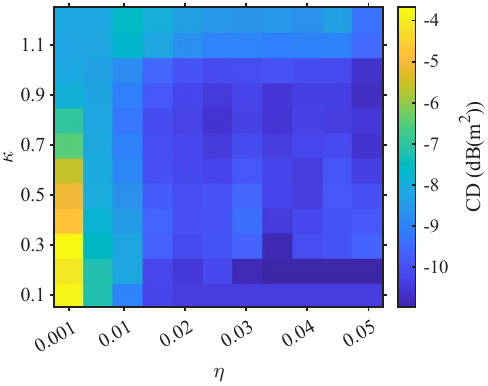}
		\label{fig:kappa_eta_sweep_c}}
	\hfill
	\subfloat[]{
		\includegraphics[width=0.23\textwidth]{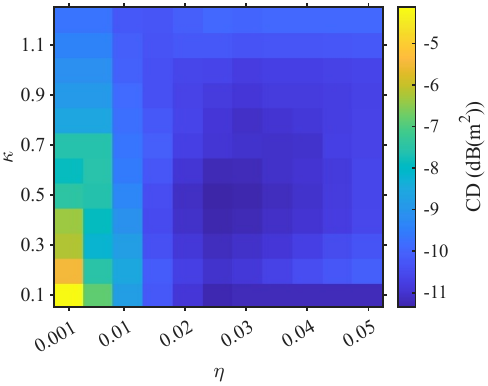}
		\label{fig:kappa_eta_sweep_d}}
	\caption{Parameter sweep results for $\kappa$ and $\eta$. (a) Averaged CD over the three canonical targets. (b) Circle. (c) Square. (d) Equilateral triangle.}
	\label{fig:kappa_eta_sweep}
\end{figure}

As described in Section~\ref{section-3}, the parameter $\kappa$ determines the radial kernel scale through $\sigma_R=\kappa\Delta R$, while $\eta$ determines the migration scale through $\delta_{\mathrm{move}}=\eta D$.

To determine a unified parameter pair, a grid search is performed over candidate values of $\kappa$ and $\eta$. For each parameter pair $(\kappa,\eta)$, the proposed MVLA-GR method is applied to reconstruct the three canonical targets. During post-processing, the final binary reconstruction is obtained by the joint thresholding rule introduced in Section~\ref{section-3}, where the retained top $p\%$ support values and the angular-support threshold $T_\theta$ are both scanned over their admissible ranges. For each target, the minimum CD obtained over all tested $(p,T_\theta)$ combinations is taken as the reconstruction error associated with the current parameter pair $(\kappa,\eta)$.

Let $\mathrm{CD}_q(\kappa,\eta)$ denote this minimum CD obtained on the $q$-th target, where
\[
q \in \{\text{circle},\text{square},\text{triangle}\}.
\]
The aggregated performance of a parameter pair is defined as the average CD over the three targets:
\begin{equation}
	J(\kappa,\eta)
	=
	\frac{1}{3}\sum_{q=1}^{3}\mathrm{CD}_q(\kappa,\eta).
	\label{eq:param_selection_obj}
\end{equation}
The final parameter pair is selected as
\begin{equation}
	(\kappa^\star,\eta^\star)
	=
	\arg\min_{\kappa,\eta} J(\kappa,\eta),
	\label{eq:param_selection_argmin}
\end{equation}
and is then fixed for all subsequent simulations and measurements in this section.

\begin{figure}[!b]
	\centering
	\includegraphics[width=0.7\columnwidth]{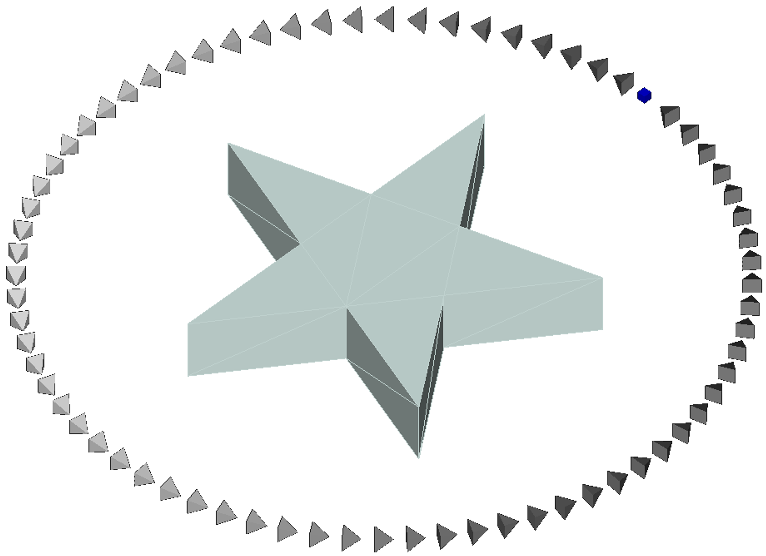}
	\caption{Ray-tracing simulation setup for the star-shaped target.}
	\label{fig:rt_scenario}
\end{figure}

Fig.~\ref{fig:kappa_eta_sweep} shows the parameter sweep results. Fig.~\ref{fig:kappa_eta_sweep}(a) presents the averaged CD map defined in~\eqref{eq:param_selection_obj}, while Fig.~\ref{fig:kappa_eta_sweep}(b)--(d) show the corresponding CD maps for the circle, square, and triangle, respectively. The minimum of the averaged CD map is attained at $\kappa^\star=0.5$ and $\eta^\star=0.04$.

\subsection{Comparative Results on the Star-Shaped Target}
\label{subsec:star_comparison}

\begin{figure*}[!t]
	\centering
	\subfloat[\footnotesize Optimal, SNR=0 dB]{\includegraphics[width=0.19\textwidth]{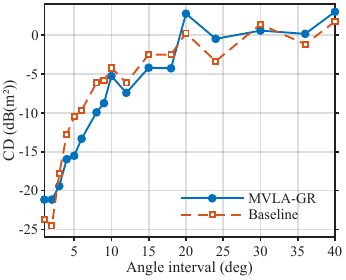}}
	\hfill
	\subfloat[\footnotesize Optimal, SNR=5 dB]{\includegraphics[width=0.19\textwidth]{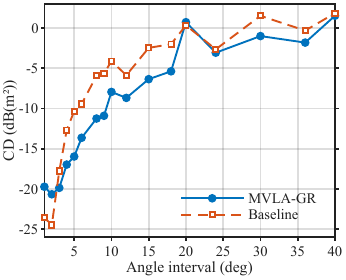}}
	\hfill
	\subfloat[\footnotesize Optimal, SNR=10 dB]{\includegraphics[width=0.19\textwidth]{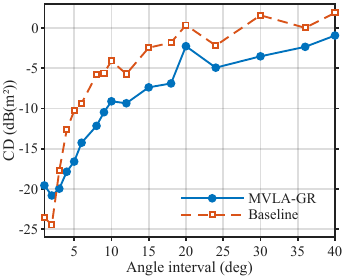}}
	\hfill
	\subfloat[\footnotesize Optimal, SNR=15 dB]{\includegraphics[width=0.19\textwidth]{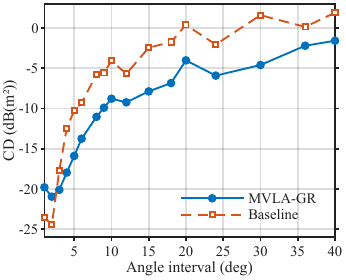}}
	\hfill
	\subfloat[\footnotesize Optimal, SNR=20 dB]{\includegraphics[width=0.19\textwidth]{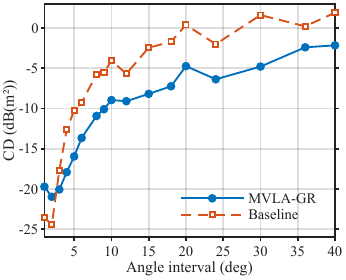}}
	
	\vspace{0.1mm}
	
	\subfloat[\footnotesize Manual, SNR=0 dB]{\includegraphics[width=0.19\textwidth]{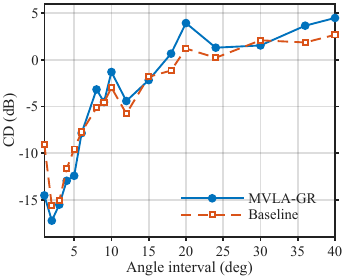}}
	\hfill
	\subfloat[\footnotesize Manual, SNR=5 dB]{\includegraphics[width=0.19\textwidth]{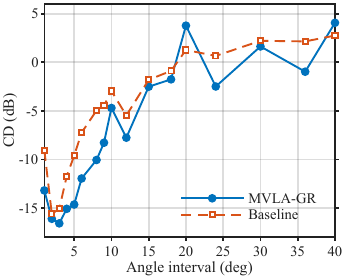}}
	\hfill
	\subfloat[\footnotesize Manual, SNR=10 dB]{\includegraphics[width=0.19\textwidth]{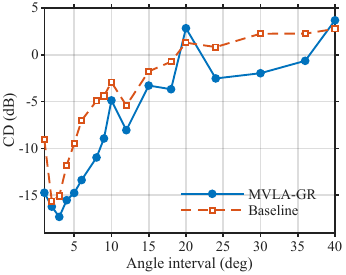}}
	\hfill
	\subfloat[\footnotesize Manual, SNR=15 dB]{\includegraphics[width=0.19\textwidth]{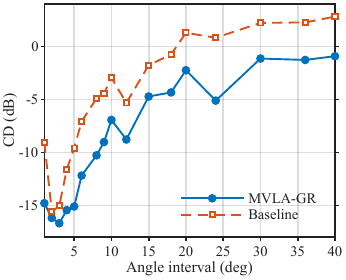}}
	\hfill
	\subfloat[\footnotesize Manual, SNR=20 dB]{\includegraphics[width=0.19\textwidth]{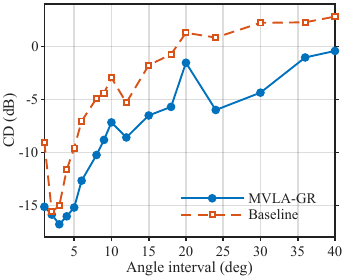}}
	\caption{CD versus angular interval. Top row: results under optimal threshold selection. Bottom row: results under calibrated manual thresholds.}
	\label{fig:cd_vs_step_combined}
\end{figure*}

With the parameter pair $(\kappa^\star,\eta^\star)=(0.5,0.04)$ fixed according to Section~\ref{subsec:param_selection}, we next evaluate the proposed MVLA-GR method on a more complex star-shaped target. Compared with the canonical targets used for parameter determination, the star-shaped target contains both sharp corners and concave structures, and therefore provides a more challenging geometry for reconstruction. The simulation configuration of the star-shaped target and the corresponding observation positions are illustrated in Fig.~\ref{fig:rt_scenario}.

To benchmark the proposed method, we consider a standard incoherent backprojection (BP) baseline. For each voxel $\mathbf{p}_n$, the baseline accumulates PDP samples at the corresponding round-trip delay across all observation views, i.e.,
\begin{equation}
	I_n^{\mathrm{BP}} = \sum_{m=1}^{M} P_m(\tau_{m,n}),
	\label{eq:bp_baseline}
\end{equation}
where $\tau_{m,n} = 2 r_{m,n}/c$ is the round-trip delay between voxel $\mathbf{p}_n$ and the $m$-th observation position. The final binary reconstruction is obtained by retaining the top-$p\%$ voxels of $\{I_n^{\mathrm{BP}}\}$. By contrast, for the proposed MVLA-GR method, the final binary reconstruction is obtained through the joint thresholding rule introduced in Section~\ref{section-3}.

Unless otherwise specified, all simulations in this subsection are conducted under the same carrier frequency, bandwidth, trajectory radius, and voxel resolution as those used in Section~\ref{subsec:param_selection}. The original ray-tracing data are generated on a dense $1^\circ$ angular grid. For a given angular interval $\Delta\theta$, the corresponding observation set is obtained by uniformly subsampling this dense grid. To reduce the dependence on the starting observation angle, the final reported CD is obtained by averaging the results over all valid angular offset patterns.

\subsubsection{Comparison Under Optimal Threshold Selection}

We first compare the proposed method with the baseline under the optimal threshold setting for each configuration. For the proposed MVLA-GR method, the post-processing parameters, namely the retained top $p\%$ support values and the angular-support threshold $T_\theta$, are jointly scanned, and the minimum CD over all tested combinations is used as the final result. For the baseline method, the top-$p\%$ threshold is scanned, and the minimum resulting CD is reported.

Fig.~\ref{fig:cd_vs_step_combined}(a)--(e) shows the reconstruction error as a function of the angular interval under five representative SNR values. In general, the CD increases as the angular interval becomes larger, since fewer observation views are available and the geometric constraints become weaker. Under most tested conditions, the proposed MVLA-GR method achieves lower CD values than the baseline, and the advantage becomes more evident as the angular sampling becomes sparser.

The SNR sensitivity of the proposed method depends strongly on the angular sampling density. When the angular interval is very small, the proposed method is relatively insensitive to SNR, whereas the influence of SNR becomes increasingly visible as the angular interval grows. This indicates that the effects of view density and observation quality are strongly coupled. When the observation views are sufficiently dense, the multi-view geometric constraints are highly redundant, and the accumulated likelihood remains well concentrated around the true contour even if the front-end peak extraction is moderately degraded by noise. As a result, the reconstruction accuracy of the proposed method varies only slightly with SNR in the dense-view regime. In contrast, when the angular interval becomes larger, the number of available views decreases and the geometric constraints become weaker, so the quality of the extracted peaks plays a more critical role, making the reconstruction performance more sensitive to SNR.

Under extremely dense angular sampling, the baseline can become competitive with the proposed method. This is because, in this regime, the observation views are already highly redundant, and the standard backprojection can directly benefit from dense multi-view accumulation. By contrast, the proposed method adopts a unified parameter setting fixed in Section~\ref{subsec:param_selection} rather than being re-optimized for this dense-view regime, which may make this unified setting slightly suboptimal when the angular sampling is extremely dense. Nevertheless, as the angular interval increases, the proposed method exhibits a clearer advantage and maintains better reconstruction accuracy under more practically relevant sparse-view conditions.

The baseline appears less sensitive to SNR than the proposed method, because the two methods are affected by noise in different ways. The baseline directly accumulates PDP samples at the corresponding round-trip delays, so noise is partially averaged out during integration. By contrast, MVLA-GR relies on peak extraction before accumulation, so noise additionally affects the front-end observation generation through missed peaks, false peaks, and peak-location perturbations, leading to a clearer dependence on SNR especially when the angular sampling is not sufficiently dense.

\begin{figure*}[!t]
	\centering
	\subfloat[\footnotesize $\Delta\theta=1^\circ$, SNR = 20 dB]{
		\includegraphics[width=0.19\textwidth]{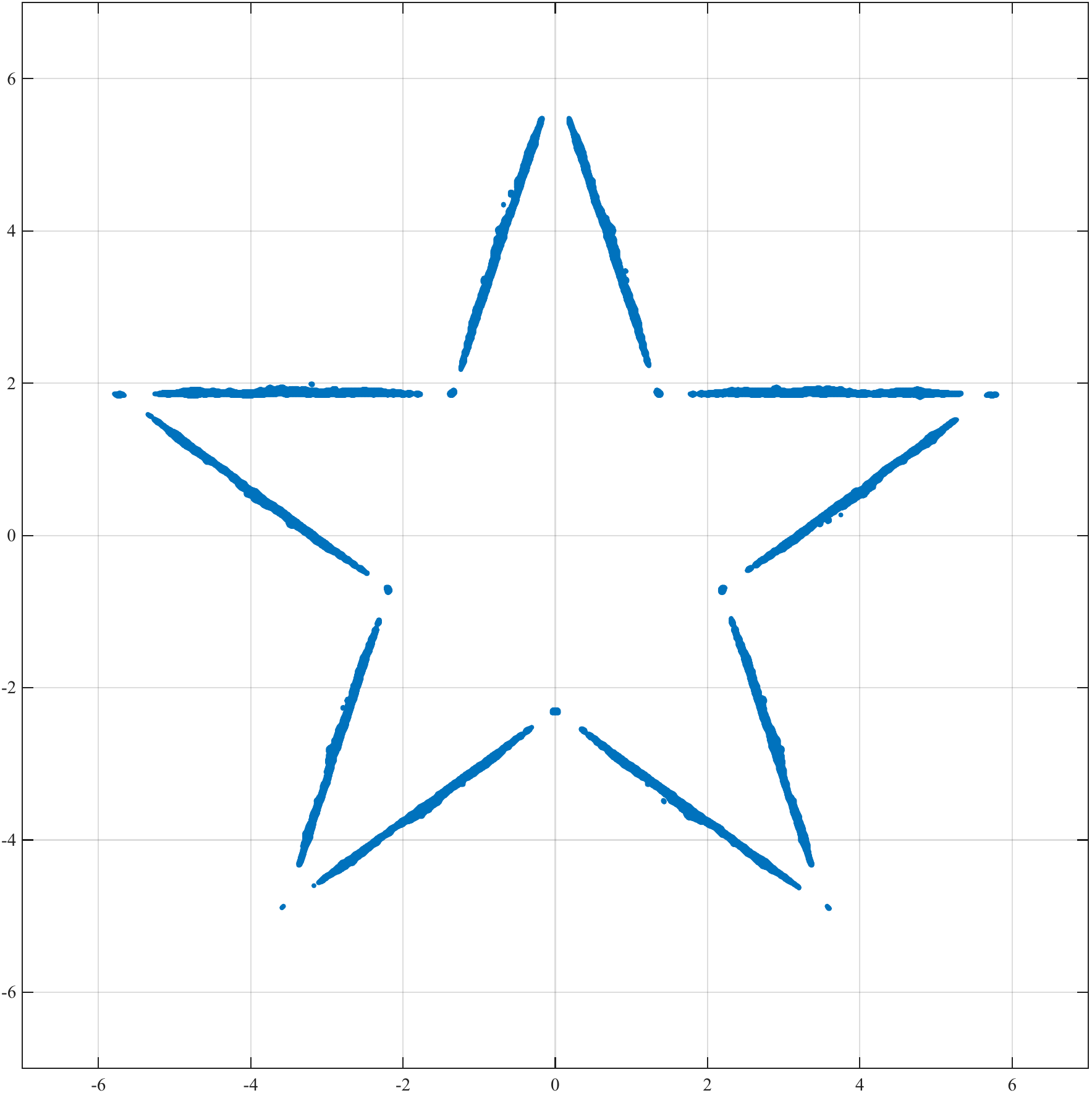}
		\label{fig:mvla_examples_a}}
	\hfill
	\subfloat[\footnotesize $\Delta\theta=5^\circ$, SNR = 20 dB]{
		\includegraphics[width=0.19\textwidth]{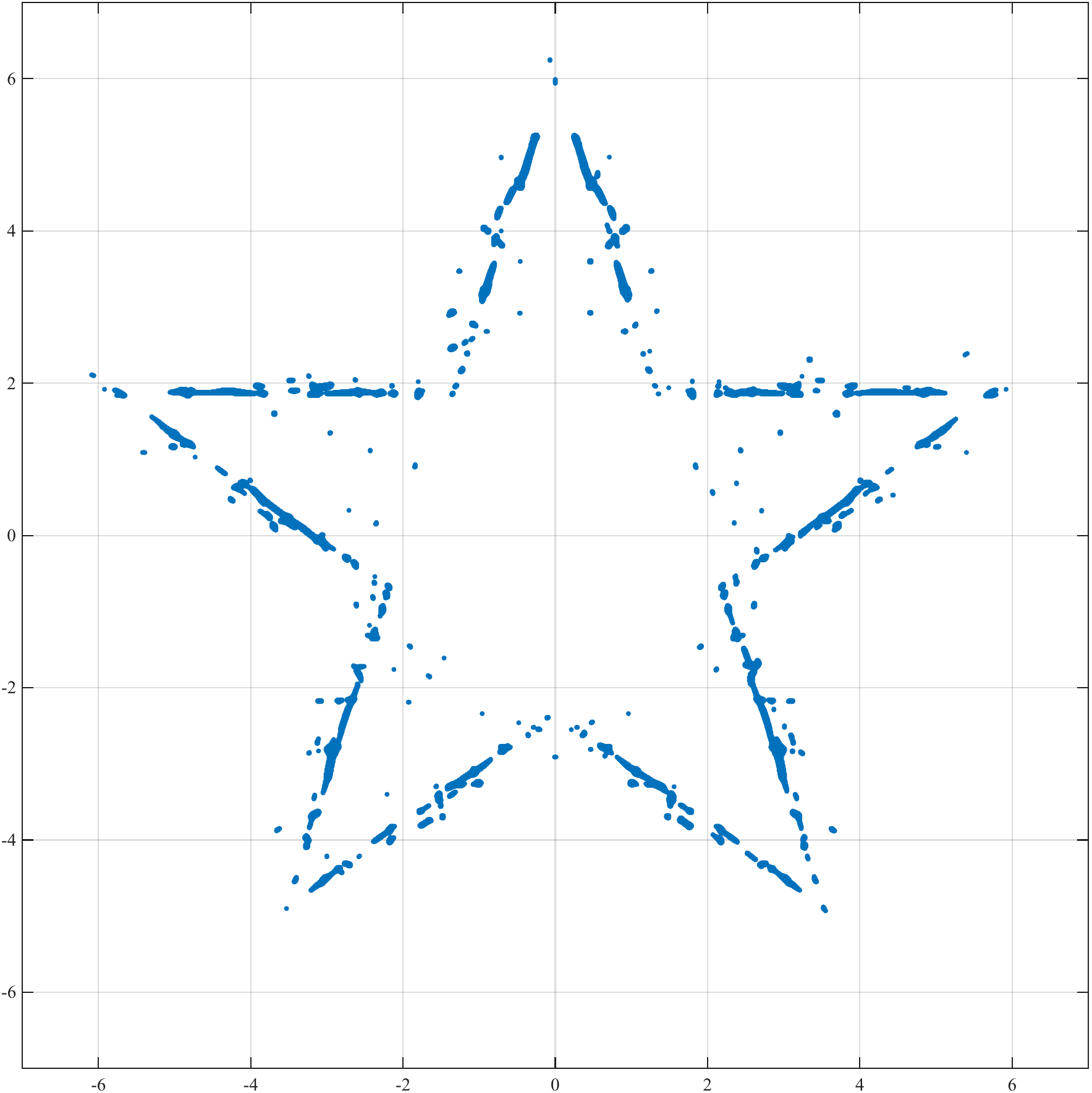}
		\label{fig:mvla_examples_b}}
	\hfill
	\subfloat[\footnotesize $\Delta\theta=10^\circ$, SNR = 20 dB]{
		\includegraphics[width=0.19\textwidth]{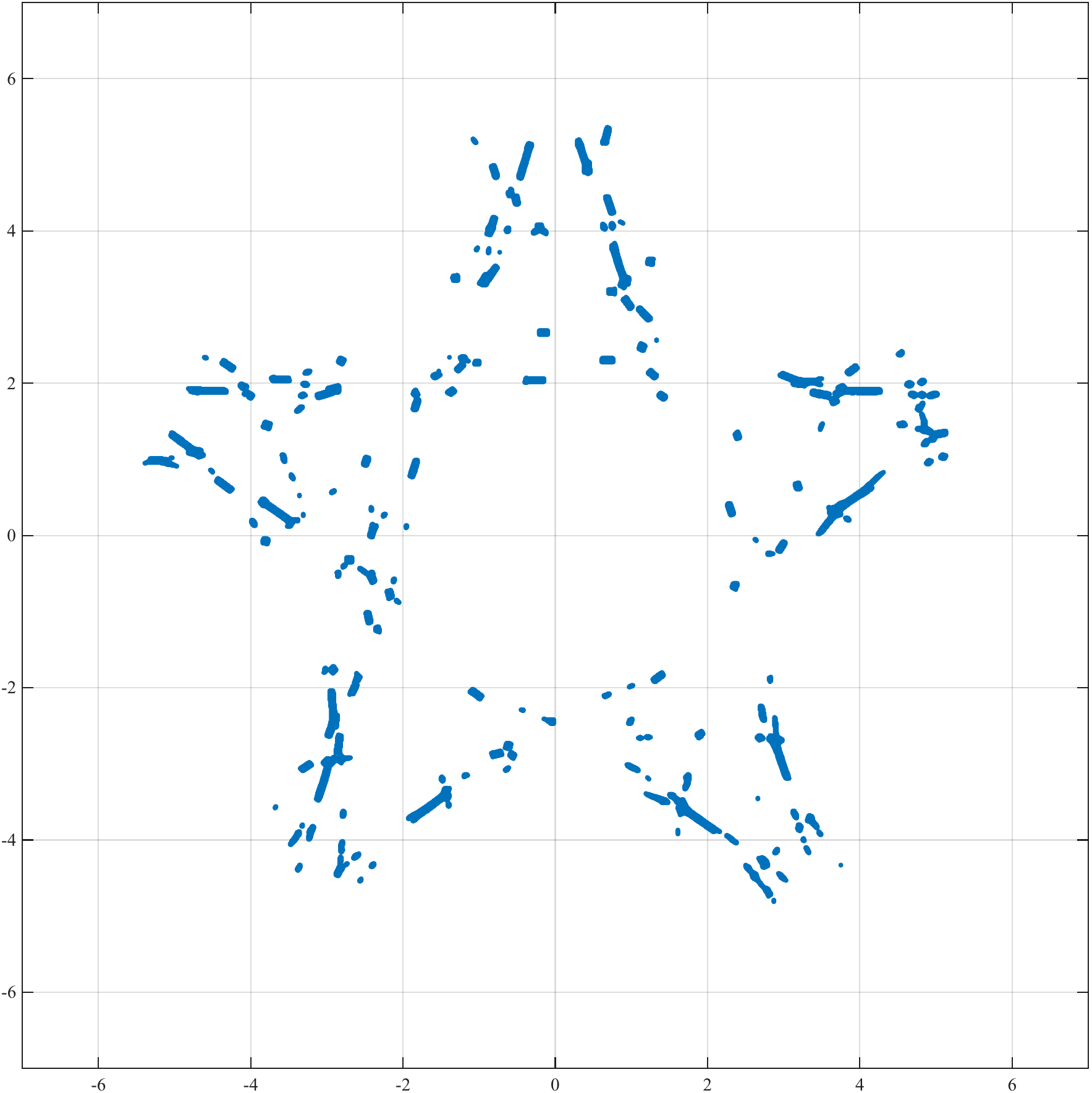}
		\label{fig:mvla_examples_c}}
	\hfill
	\subfloat[\footnotesize $\Delta\theta=30^\circ$, SNR = 20 dB]{
		\includegraphics[width=0.19\textwidth]{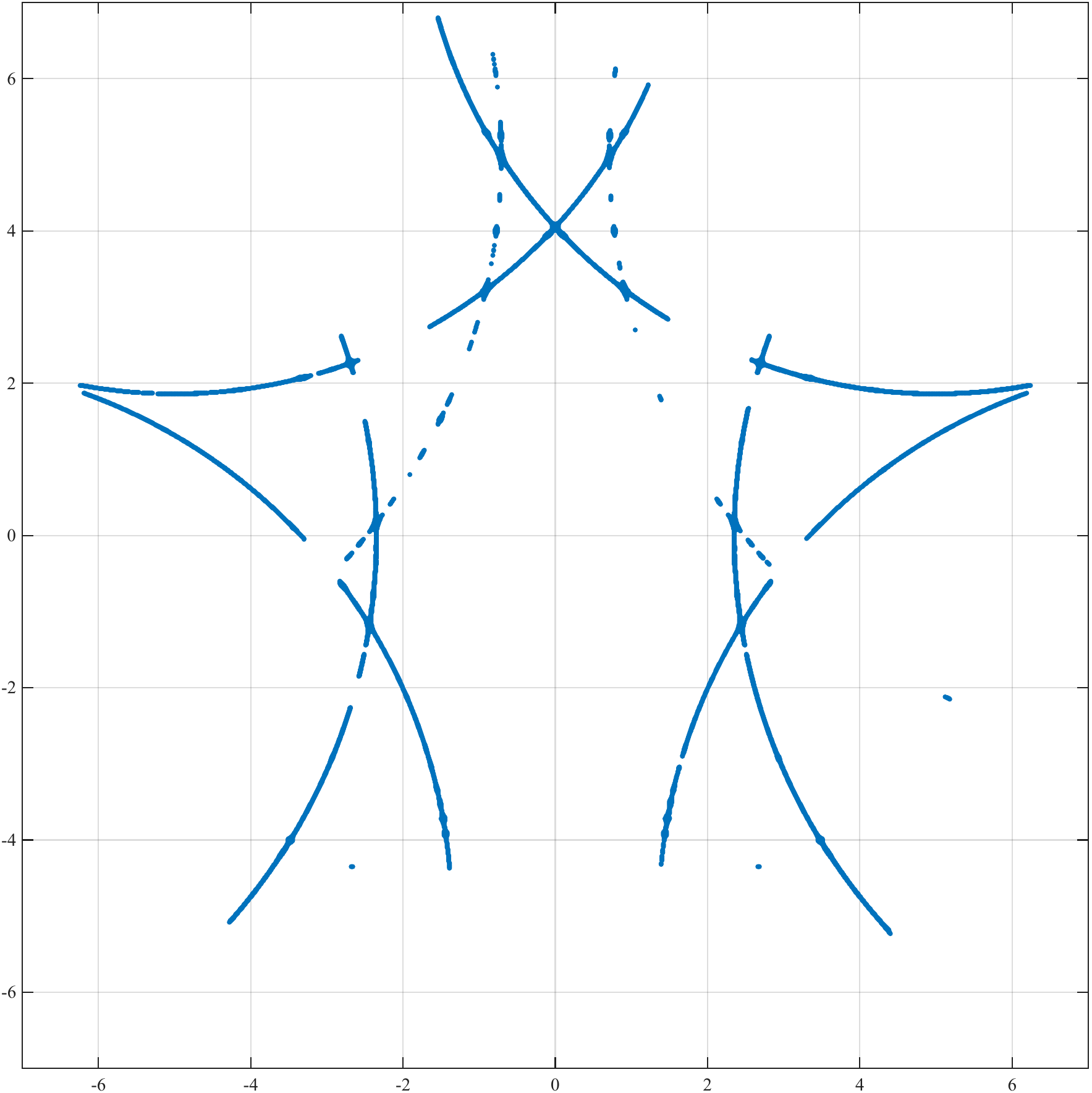}
		\label{fig:mvla_examples_d}}
	
	\vspace{0.1mm}
	
	\subfloat[\footnotesize $\Delta\theta=5^\circ$, SNR = 15 dB]{
		\includegraphics[width=0.19\textwidth]{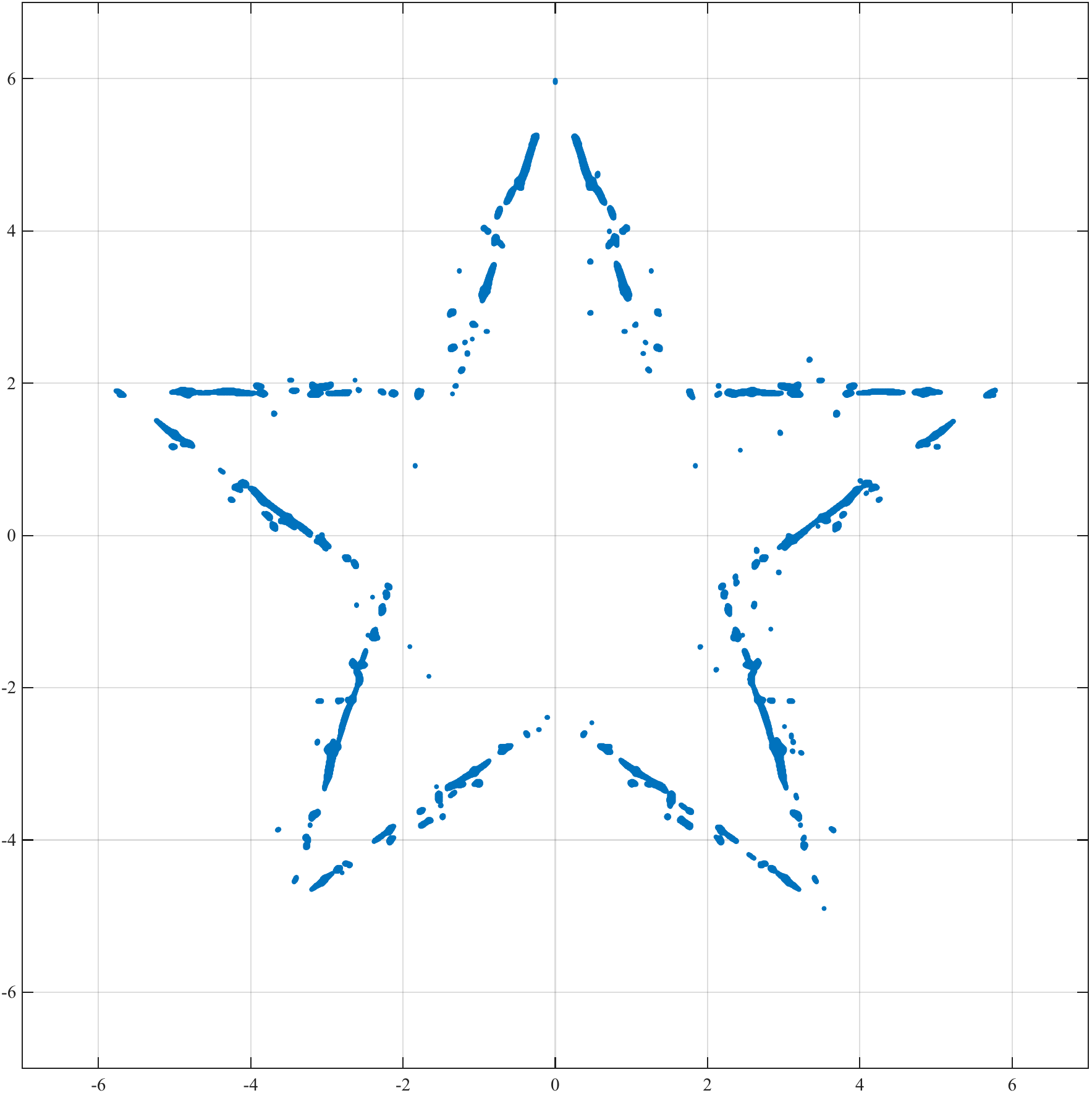}
		\label{fig:mvla_examples_e}}
	\hfill
	\subfloat[\footnotesize $\Delta\theta=5^\circ$, SNR = 10 dB]{
		\includegraphics[width=0.19\textwidth]{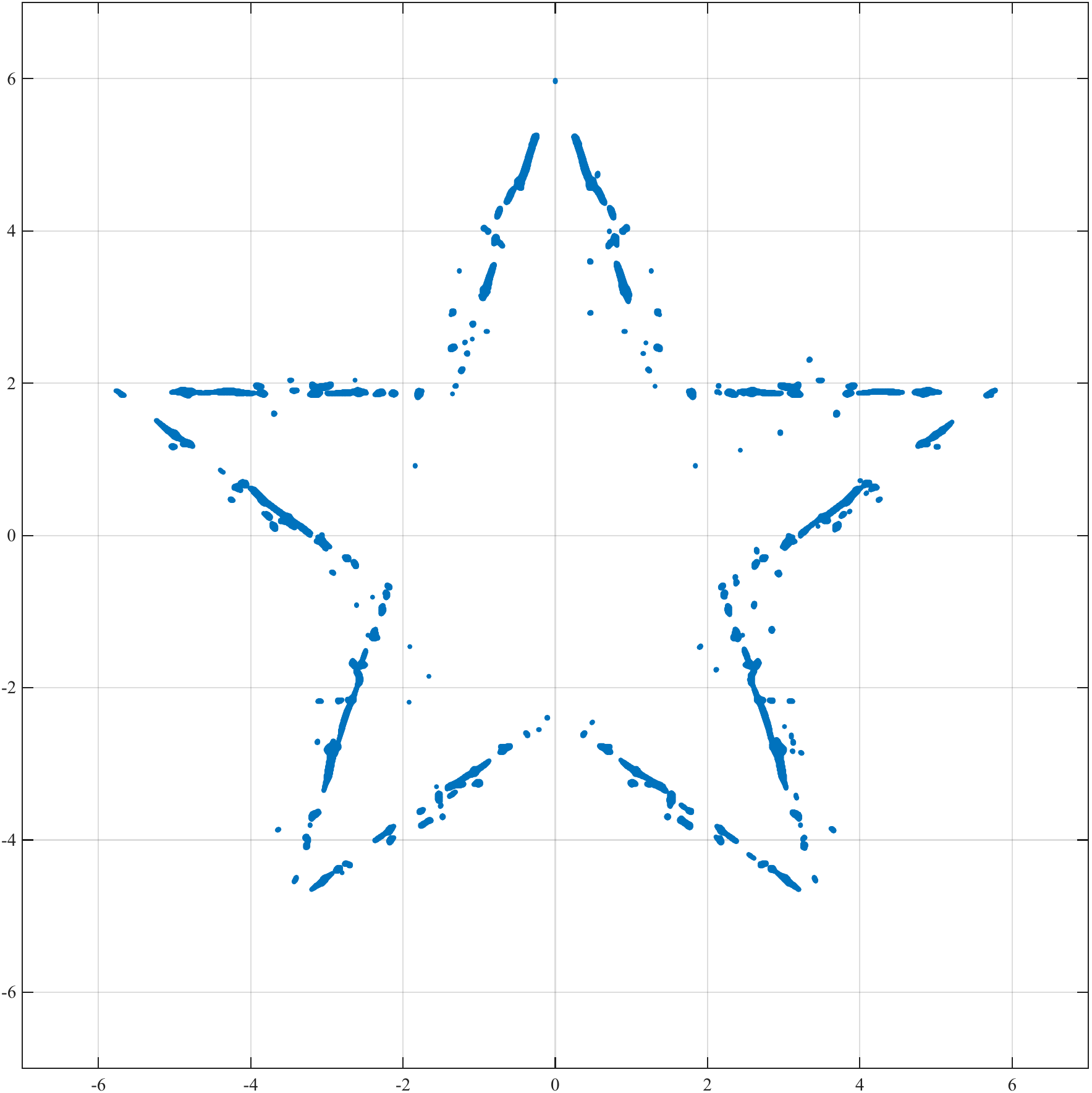}
		\label{fig:mvla_examples_f}}
	\hfill
	\subfloat[\footnotesize $\Delta\theta=5^\circ$, SNR = 5 dB]{
		\includegraphics[width=0.19\textwidth]{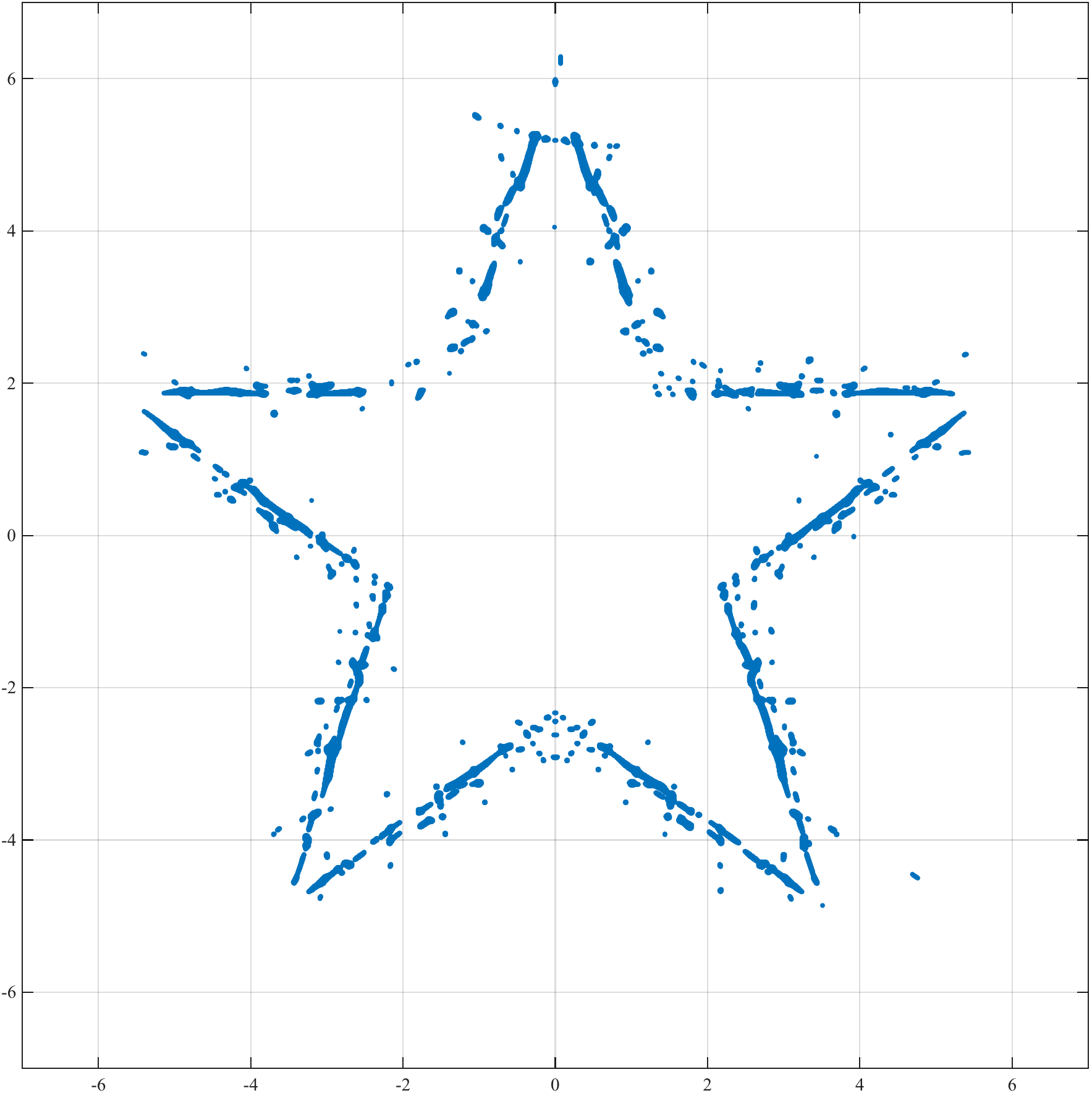}
		\label{fig:mvla_examples_g}}
	\hfill
	\subfloat[\footnotesize $\Delta\theta=5^\circ$, SNR = 0 dB]{
		\includegraphics[width=0.19\textwidth]{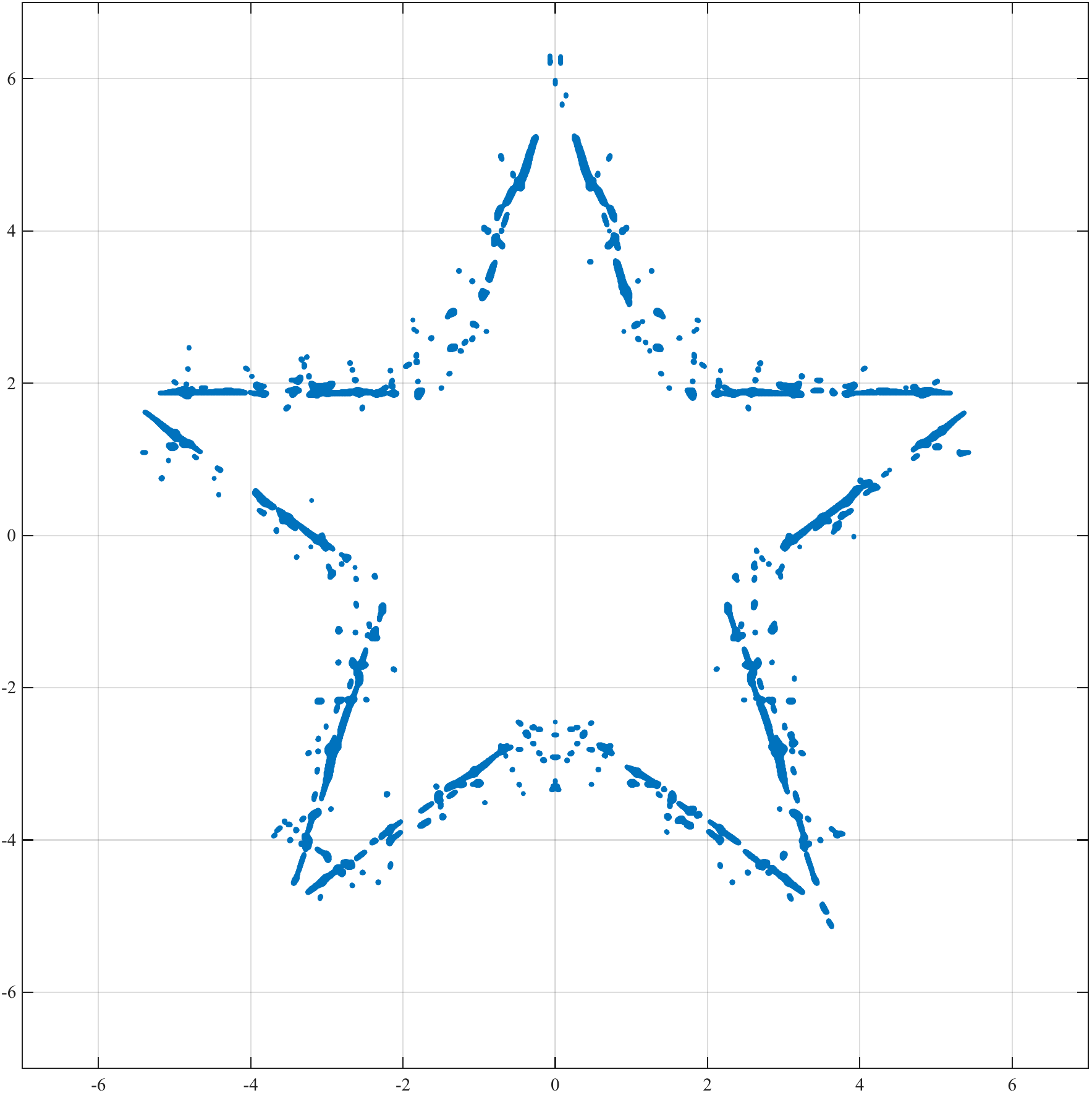}
		\label{fig:mvla_examples_h}}
	\caption{Representative reconstruction results of the proposed MVLA-GR method under different sensing conditions.}
	\label{fig:mvla_examples}
\end{figure*}

Representative reconstruction results of the proposed MVLA-GR method are shown in Fig.~\ref{fig:mvla_examples}. The first row illustrates the effect of angular interval when the SNR is fixed at 20~dB, while the second row shows the effect of SNR when the angular interval is fixed at $5^\circ$. These visual results are consistent with the quantitative CD curves: denser angular sampling and higher SNR both lead to clearer and more complete reconstruction of the star-shaped contour.

\subsubsection{Comparison Under Calibrated Manual Thresholds}

Although the optimal-threshold results reveal the best achievable reconstruction performance under each configuration, such exhaustive tuning is generally unavailable in practice. We therefore further consider a calibrated manual-threshold setting. Specifically, for each SNR, the retained top $p\%$ support ratio and the angular-support threshold $T_\theta$ of the proposed MVLA-GR method are fixed according to the average of their optimal values over different angular intervals at that SNR. For the baseline, the top-$p\%$ threshold is calibrated in the same way. The resulting thresholds are then reused for all angular intervals under the corresponding SNR without further tuning. The calibrated threshold values used for each SNR are summarized in Table~\ref{tab:manual_thresholds}.

\begin{table}[!h]
	\caption{Calibrated Manual Thresholds}
	\label{tab:manual_thresholds}
	\centering
	\begin{tabular}{cccc}
		\hline
		\textbf{SNR (dB)} & \textbf{MVLA-GR: $p$} & \textbf{MVLA-GR: $T_\theta$ (deg)} & \textbf{BP: $p$} \\
		\hline
		0  & 0.6090 & 30.4115 & 0.3000 \\
		5  & 0.6374 & 52.5926 & 0.2975 \\
		10 & 0.6814 & 69.9835 & 0.3008 \\
		15 & 0.6736 & 84.2016 & 0.2938 \\
		20 & 0.7008 & 96.8477 & 0.2942 \\
		\hline
	\end{tabular}
\end{table}

Fig.~\ref{fig:cd_vs_step_combined}(f)--(j) shows the corresponding CD curves under this calibrated manual-threshold setting. Compared with the optimal-threshold results in Fig.~\ref{fig:cd_vs_step_combined}(a)--(e), both methods exhibit performance degradation, which is expected since the thresholds are no longer individually optimized for each configuration. However, the relative performance trend between the two methods remains informative. In particular, as the SNR increases, the proposed MVLA-GR method shows a clearer advantage over the baseline under most angular intervals. This indicates that, although the fixed thresholds introduce some mismatch, the proposed method is still able to exploit multi-view geometric consistency more effectively when the observation quality is sufficiently good.

Table~\ref{tab:manual_thresholds} also shows that the calibrated thresholds of the proposed MVLA-GR method exhibit a clear SNR dependence: the retained top-$p\%$ ratio increases with SNR because cleaner observations contain fewer artifact responses, and $T_\theta$ increases monotonically with SNR because more complete and reliable peaks at higher SNR make true scattering voxels remain continuously supported over a wider angular range. By contrast, the calibrated top-$p\%$ threshold of the baseline remains relatively stable across SNR, which is consistent with its low SNR sensitivity discussed above.

The SNR-dependent advantage of MVLA-GR can be explained from two perspectives. When the SNR is moderate or high, the extracted peaks are sufficiently reliable, and the geometric-consistency mechanism of MVLA-GR can still be effectively exploited even under fixed thresholds, so its higher reconstruction ceiling translates into a clear advantage over the baseline. When the SNR is low, peak extraction becomes more vulnerable to missed peaks, false peaks, and location perturbations, which makes MVLA-GR more sensitive to threshold mismatch. Overall, the comparison across Fig.~\ref{fig:cd_vs_step_combined} suggests that MVLA-GR not only achieves a higher best-case performance, but also preserves its advantage under calibrated manual thresholds when the SNR is moderate or high, indicating meaningful robustness in practical thresholding conditions.

\subsection{Vehicle Measurement Validation}
\label{subsec:measurement_validation}

To further assess the proposed MVLA-GR framework under practical conditions, real measurement data collected from a vehicle target are considered in this subsection.

\begin{figure}[!b]
	\centering
	\includegraphics[width=0.7\columnwidth]{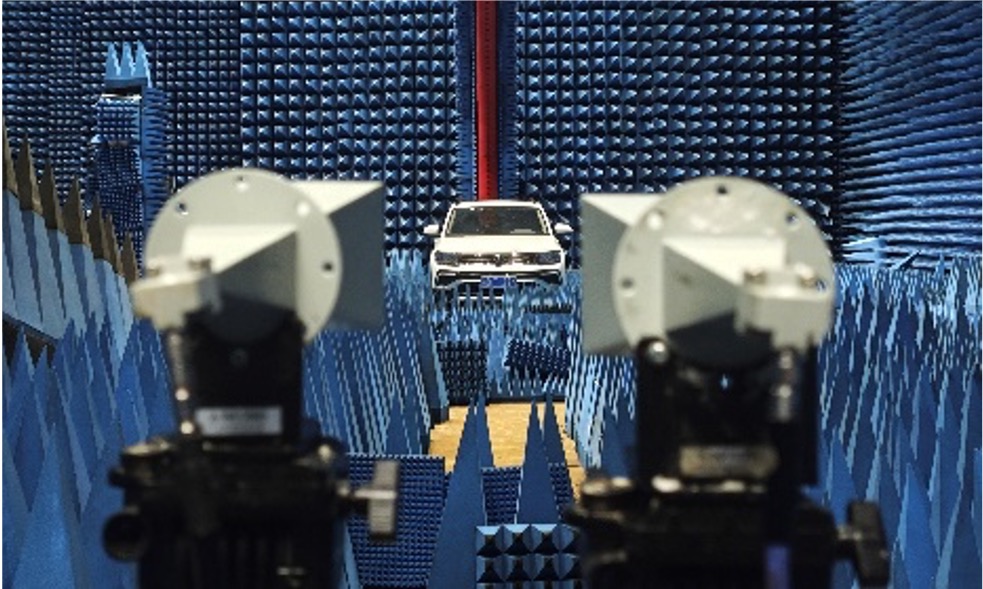}
	\caption{Measurement setup in the anechoic chamber.}
	\label{fig:meas_scenario}
\end{figure}

\subsubsection{Measurement Setup}

The measurement is conducted in an anechoic chamber to suppress external interference and unwanted environmental reflections. The measurement platform follows the setup reported in~\cite{zhang2026rcs}, where its calibration accuracy has been verified using a metal sphere as a reference target. A Volkswagen T-CROSS vehicle is used as the target, whose physical dimensions are approximately $4.22 \times 1.76 \times 1.60~\mathrm{m}$ (length $\times$ width $\times$ height).

The measurement system consists of a vector network analyzer (VNA), two horn antennas, and the associated RF front-end components. The two antennas are placed in close proximity to emulate a quasi-monostatic sensing configuration. During the experiment, the vehicle is mounted on a turntable, while the antennas remain fixed and point toward the target, as shown in Fig.~\ref{fig:meas_scenario}. By rotating the turntable, the target is observed from multiple aspect angles, which is equivalent to collecting multi-view measurements around the target. The main measurement configuration is summarized in Table~\ref{tab:measurement_configuration}.

\begin{table}[!t]
	\caption{Measurement Configuration}
	\label{tab:measurement_configuration}
	\centering
	\begin{tabular}{cc}
		\hline
		\textbf{Parameters} & \textbf{Value / Type} \\
		\hline
		Carrier frequency $f_c$ / Bandwidth $B$ & 36~GHz / 3~GHz \\
		Range resolution $\Delta R$ & 0.05~m \\
		Number of views $M$ ($\Delta\theta$) & 36 ($10^\circ$) \\
		Target--antenna distance & 12~m \\
		Antenna height & 0.9~m \\
		Antenna type (Beamwidth) & Horn antenna ($14.39^\circ$) \\
		Target size & $4.22 \times 1.76 \times 1.60~\mathrm{m}$ \\
		\hline
	\end{tabular}
\end{table}

\subsubsection{Measured Data Processing and Reconstruction Configuration}

For each observation angle, the VNA records the complex frequency response over the operating bandwidth. After calibration, the corresponding CIR is obtained by inverse Fourier transform, and the PDP is computed as the squared magnitude of the CIR. Since the measured data already contain practical noise and system imperfections, no additional synthetic noise is introduced in this experiment. Following the same phase-free reconstruction framework used in the simulation part, dominant multipath components are extracted from the measured PDPs and represented by their peak distances and peak powers.

The extracted peaks are then fed into the proposed MVLA-GR method. The model parameters are directly inherited from Section~\ref{subsec:param_selection}, namely $\kappa^\star = 0.5$ and $\eta^\star = 0.04$. Since the antennas have a finite beamwidth rather than omnidirectional radiation patterns, each distance observation is only projected within the corresponding beam-covered angular sector. This beam constraint is incorporated into the likelihood accumulation process to suppress support outside the illuminated region.

The reconstruction region is discretized with a voxel size of 0.005~m. After likelihood accumulation, the final binary result is obtained using manually selected threshold parameters for visualization. Quantitative CD evaluation is omitted here, since accurate voxel-wise ground-truth occupancy labels are not available for the real vehicle target. The reconstruction result is therefore assessed visually.

\begin{figure}[!t]
	\centering
	\subfloat[]{
		\includegraphics[width=0.23\textwidth]{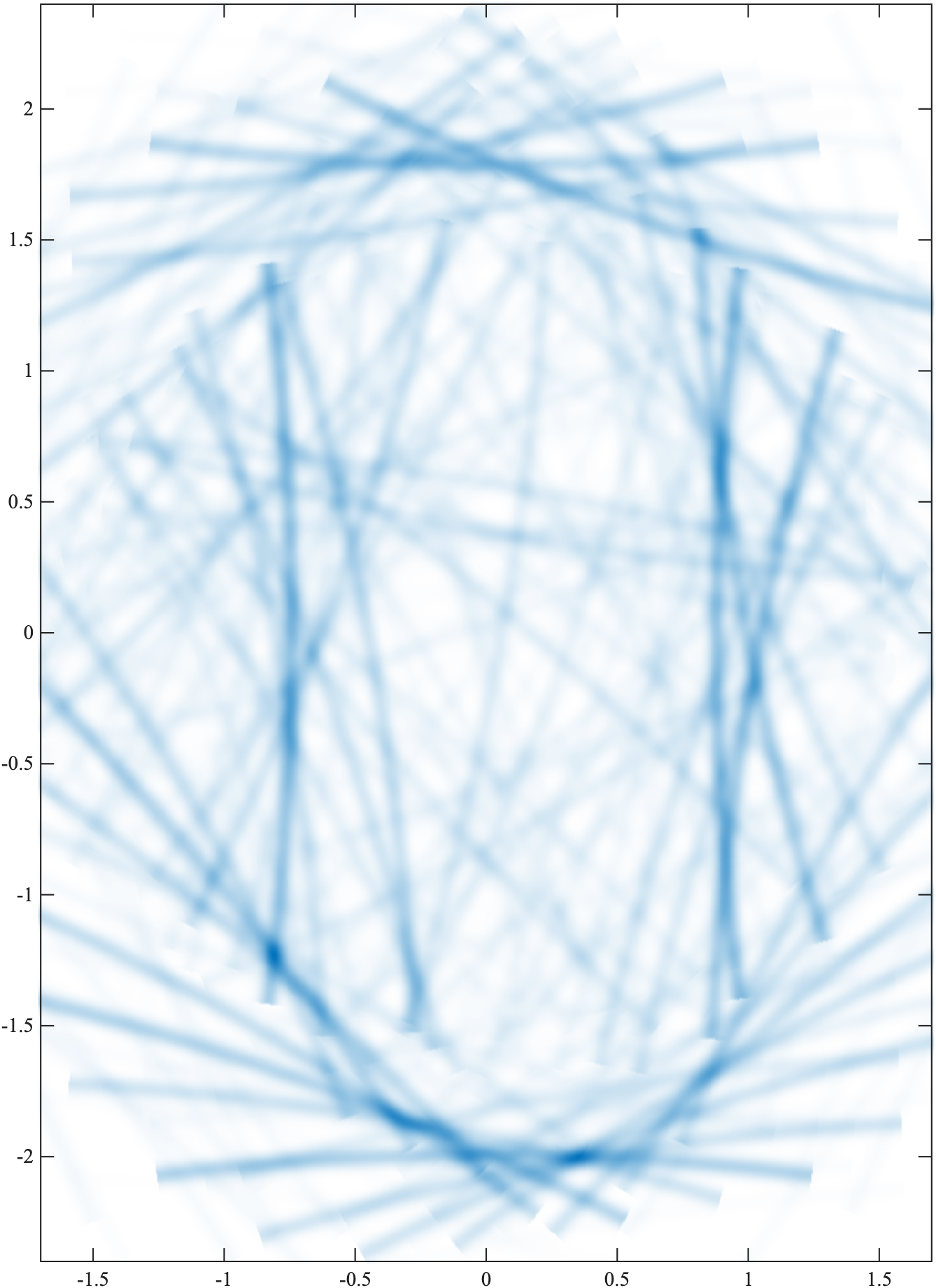}
		\label{fig:vehicle_meas_support}}
	\hfill
	\subfloat[]{
		\includegraphics[width=0.23\textwidth]{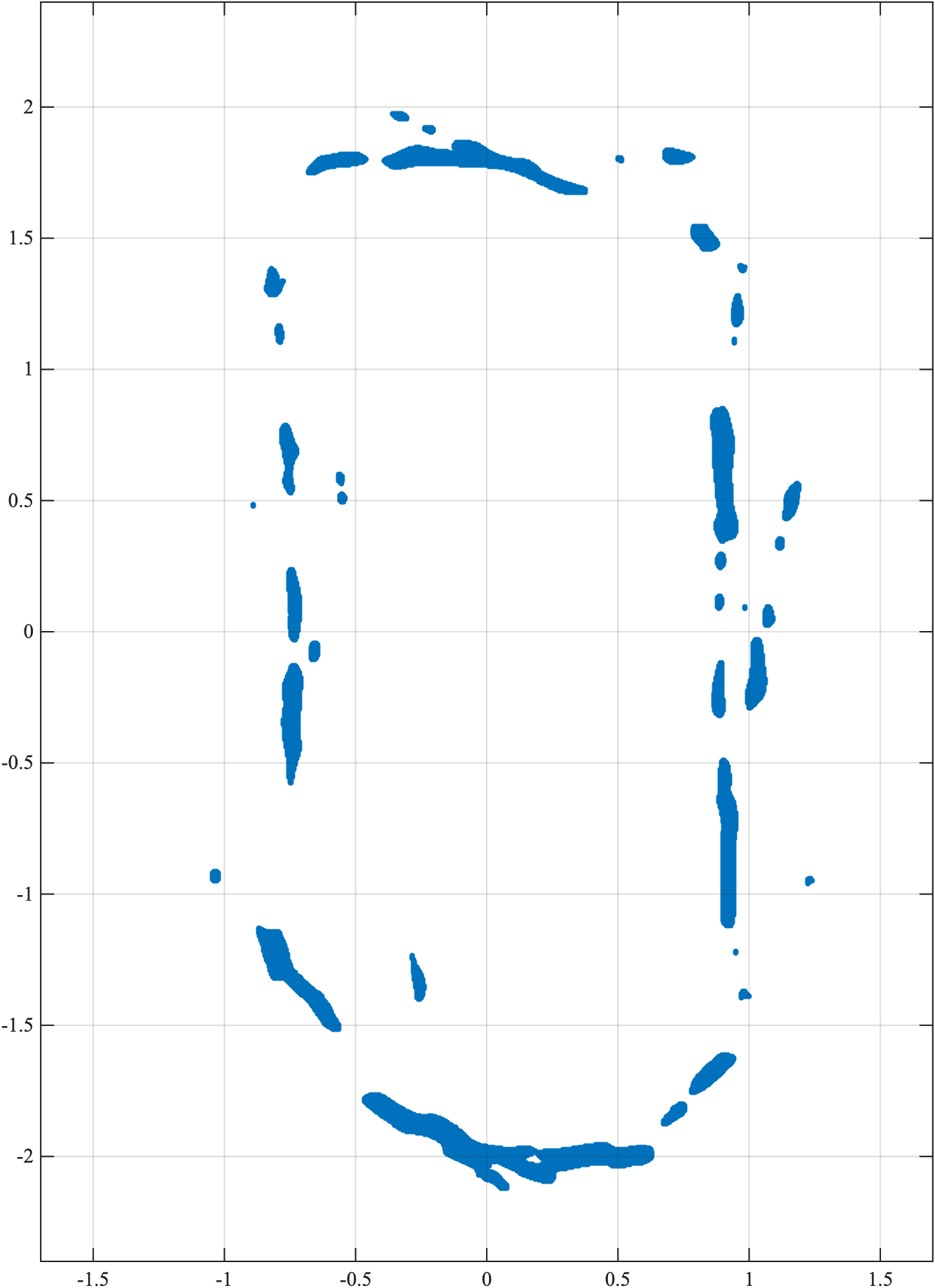}
		\label{fig:vehicle_meas_binary}}
	\caption{Measured reconstruction results of the vehicle target obtained by the proposed MVLA-GR method. (a) Continuous support map before thresholding. (b) Final binary reconstruction result after thresholding.}
	\label{fig:vehicle_measurement_result}
\end{figure}

\subsubsection{Measurement Results}

The measured reconstruction results of the vehicle target are shown in Fig.~\ref{fig:vehicle_measurement_result}. Fig.~\ref{fig:vehicle_measurement_result}(a) presents the continuous support map obtained after multi-view likelihood accumulation, while Fig.~\ref{fig:vehicle_measurement_result}(b) shows the final binary reconstruction result after thresholding.

It can be seen that the continuous support map already reveals the main horizontal contour of the vehicle, although the support remains spatially diffuse before thresholding. After thresholding, the main contour becomes significantly clearer, and the elongated body structure as well as the overall aspect ratio of the vehicle are well preserved. These results indicate that the proposed likelihood accumulation mechanism remains effective under practical measurement conditions.

Compared with the idealized simulation results, the measured reconstruction is less regular and exhibits local discontinuities and some contour thickening. This is expected because the dominant measured scattering points originate from electromagnetically strong parts such as edges, corners, and metallic structures rather than from the outermost physical boundary, the 3D vehicle geometry is projected onto a horizontal imaging plane, and the measured data inevitably contain residual noise, calibration errors, and other non-ideal scattering effects. The recovered contour should therefore be interpreted as the effective projection of dominant scatterers on the imaging plane rather than the exact physical footprint of the vehicle.

Nevertheless, the reconstructed result still preserves the major geometric characteristics of the vehicle. This provides evidence that the proposed MVLA-GR framework is not limited to idealized ray-tracing data, but can also operate on real measured observations.

\section{Conclusion}
\label{section-5}

In this paper, we addressed the problem of phase-free environmental geometry reconstruction in ISAC systems, where the absolute phase reference of each CIR measurement and its consistency across observation positions are challenging to maintain, due to both hardware imperfections such as oscillator drift and timing offsets, and physical-layer factors such as material- and viewpoint-dependent reflection phases. The proposed MVLA-GR method reconstructs target geometry from CIR measurements by exploiting the geometric consistency of multi-view propagation paths, using only delay and power information extracted from the power delay profile. A joint thresholding strategy combining response magnitude and angular support continuity is developed to suppress reconstruction artifacts and convert the continuous support map into a binary geometry estimate.

Simulation results on canonical geometric targets and a complex star-shaped target demonstrated that the proposed method achieves lower reconstruction error than the incoherent BP baseline in the majority of tested conditions. Real-world vehicle measurements at 36 GHz further verified the practical applicability of the proposed framework, where the main geometric characteristics of the vehicle were successfully recovered from phase-free CIR observations.

Future work will further investigate the integration of MVLA-GR with ISAC tracking modules, where target trajectory information from cooperative sensing can guide the segmentation of CIR observations and enable real-time geometry reconstruction of moving targets in 6G mobile networks.

\renewcommand{\baselinestretch}{0.77}
\selectfont
\bibliographystyle{IEEEtran}
\bibliography{refs}

\end{document}